\documentclass[aps,prb,twocolumn,showpacs,superscriptaddress,floatfix]{revtex4}
\usepackage{graphicx,amsfonts,amssymb,amsmath,mathrsfs,hyperref,bm}
\begin{document}

\title{Emergence of string-valence bond solid state in the frustrated 
$J_1-J_2$ 
transverse field Ising model on the square lattice}

\author{M. Sadrzadeh}
\affiliation{Department of Physics, Sharif University of Technology, P.O.Box 11155-9161, Tehran, Iran}
\email{marzieh\_sadrzadeh@physics.sharif.edu}
\affiliation{School of Physics, Institute for Research in Fundamental Sciences (IPM), Tehran 19395-5531, Iran}

\author{R. Haghshenas}
\affiliation{Department of Physics, Sharif University of Technology, P.O.Box 11155-9161, Tehran, Iran}

\author{S. S. Jahromi}
\affiliation{Department of Physics, Sharif University of Technology, P.O.Box 11155-9161, Tehran, Iran}
\affiliation{School of Nano Sciences, Institute for Research in Fundamental Sciences (IPM), Tehran 19395-5531, Iran}

\author{A. Langari}
\affiliation{Department of Physics, Sharif University of Technology, P.O.Box 
11155-9161, Tehran, Iran}
\affiliation{School of Physics, Institute for Research in Fundamental Sciences (IPM), Tehran 19395-5531, Iran}
\affiliation{Center of excellence in Complex Systems and Condensed Matter (CSCM), Sharif University of Technology, Tehran 1458889694, Iran}
\email{langari@sharif.edu}

\begin{abstract}
We investigate the ground state nature of the transverse field Ising model on the 
$J_1-J_2$ square lattice at the highly frustrated point $J_2/J_1=0.5$.
At zero field, the model has an exponentially large degenerate 
classical ground state, which can be affected by quantum fluctuations for
non-zero field toward a unique quantum ground state.
We consider two types of quantum 
fluctuations, harmonic ones by using linear spin wave theory (LSWT) 
with single-spin flip excitations above a long range magnetically ordered 
background and
anharmonic fluctuations, by employing a cluster-operator approach (COA) with 
multi-spin cluster type fluctuations above a non-magnetic cluster ordered 
background. Our findings reveal that the harmonic fluctuations of LSWT 
fail to lift 
the extensive degeneracy as well as signaling a
violation of the Hellmann-Feynman theorem. However, the string-type anharmonic 
fluctuations of COA are able to lift 
the degeneracy toward a string-valence bond solid (VBS) state, which is 
obtained from an effective theory consistent with the Hellmann-Feynman theorem 
as well. Our results are further confirmed by implementing numerical tree tensor 
network simulation. The emergent non-magnetic 
string-VBS phase is gapped and breaks 
lattice rotational symmetry with only two-fold degeneracy, which bears a 
continuous quantum phase transition at $\Gamma/J_1 \cong 0.50$ to the quantum 
paramagnet 
phase of high fields. The critical behavior is characterized by $\nu \cong 1.0$
and $\gamma \cong 0.33$ exponents.

\end{abstract}


\maketitle
\section{Introduction}
\label{introduction}

Geometric frustration in quantum magnets results in emergence of many intriguing exotic phases of matter, ranging from resonating 
valence bond solid (VBS) phases with broken spatial symmetry to spin 
liquids with fractional quasi-particle excitations \cite{balents2010spin}. It has 
further been shown that the geometric frustration plays an important role in the 
physics of non-Fermi liquid of doped Mott insulators and high-Tc superconductors 
\cite{sheckelton2012possible,PhysRevB.85.104416,liu2015triplet,harland2016plaquette,nembrini2016tracking}. Typically, 
frustrated magnetic systems show extensive degeneracy of their ground states in 
the 
classical limit, which can be lifted by addition of thermal or quantum 
fluctuations, or perturbations such as spin-orbit interactions, spin-lattice couplings, further 
neglected exchange terms and impurities. It would lead to the emergence of 
exotic collective quantum behaviors. 

One of the simplest and hence most tractable models featuring such an interplay 
between the geometric frustration and quantum fluctuations is the 
spin-$\frac{1}{2}$ $J_1-J_2$ antiferromagnetic Heisenberg model on the square 
lattice, which is a suitable candidate for a quantum spin liquid state and is 
highly relevant to cuprates and Fe-based superconductors 
\cite{xu:2008}. It has 
already been shown that the ground state of the system in the highly frustrated 
point, $J_2/J_1=0.5$, is given by a non-magnetic state emerging as an 
intermediate phase between N\'{e}el and striped antiferromagnetic (AFM) 
states in the small and large limit of $J_2/J_1$ coupling, respectively. 
However, the true nature of the intermediate non-magnetic phase is still under 
debate. Early and recent studies have proposed different candidate ground states for the 
intermediate region around $J_2/J_1=0.5$, such as a dimer VBS with both 
translational and rotational broken symmetries \cite{singh:1999,metavitsiadis:2014}, plaquette VBS with broken translational symmetry but with 
rotational symmetry preserved 
\cite{Isaev:2009,yu:2012,doretto2014plaquette}, gapless spin liquid 
\cite{hu:2013,wang:2013,Gong:2014PRL,morita:2015}, and 
gapped $Z_2$ spin liquid phases 
\cite{jiang:2012,mezzacapo:2012,li:2012,ren2014cluster}. 

Our aim in this paper is to shed light on the true nature of this intermediate 
magnetically-disordered state by introducing both quantum fluctuations and anisotropies in the spin space to lift 
the extensive degeneracy of the classical system towards a quantum ordered ground state. 
We can introduce anisotropies to the bonds of the spin-$\frac{1}{2}$ $J_1-J_2$ 
Heisenberg model on the square lattice by breaking the $SU(2)$ symmetry and 
reducing the Heisenberg interactions of $J_1-J_2$ bonds to $XXZ$ couplings. Such spin 
anisotropy is relevant 
theoretically \cite{benyoussef1998frustrated,bishop2008effect} 
as well as experimentally \cite{yamaki2013ground,higashinaka2014pronounced,nisoli2013colloquium}. 
For the large limit of Ising anisotropies, the $XXZ$ model 
behaves equivalently to a 
transverse field Ising (TFI) model on the $J_1-J_2$ square lattie. Such TFI
model with an interplay between frustration and quantum fluctuations, can reveal what happens, by reduction of symmetry from $SU(2)$ to $Z_2$, for the true nature of the 
under-debate non-magnetic ground state of the Heisenberg model at highly 
frustrated point $J_2/J_1=0.5$.

Moreover, the 2D TFI model is a prototype frustrated magnetic model, which 
received much attention, to explore novel 
emergent phases \cite{Suzuki:book,Kalz:2009,Dutta:2010}. The 
ground state of 2D TFI model at the highly-frustrated point, to the best of our 
knowledge, is not known. It is challenging to find a ground state, which 
is a result of quantum fluctuations on an extensive degenerate ground space.

\begin{figure}
\includegraphics[width=0.9\columnwidth]{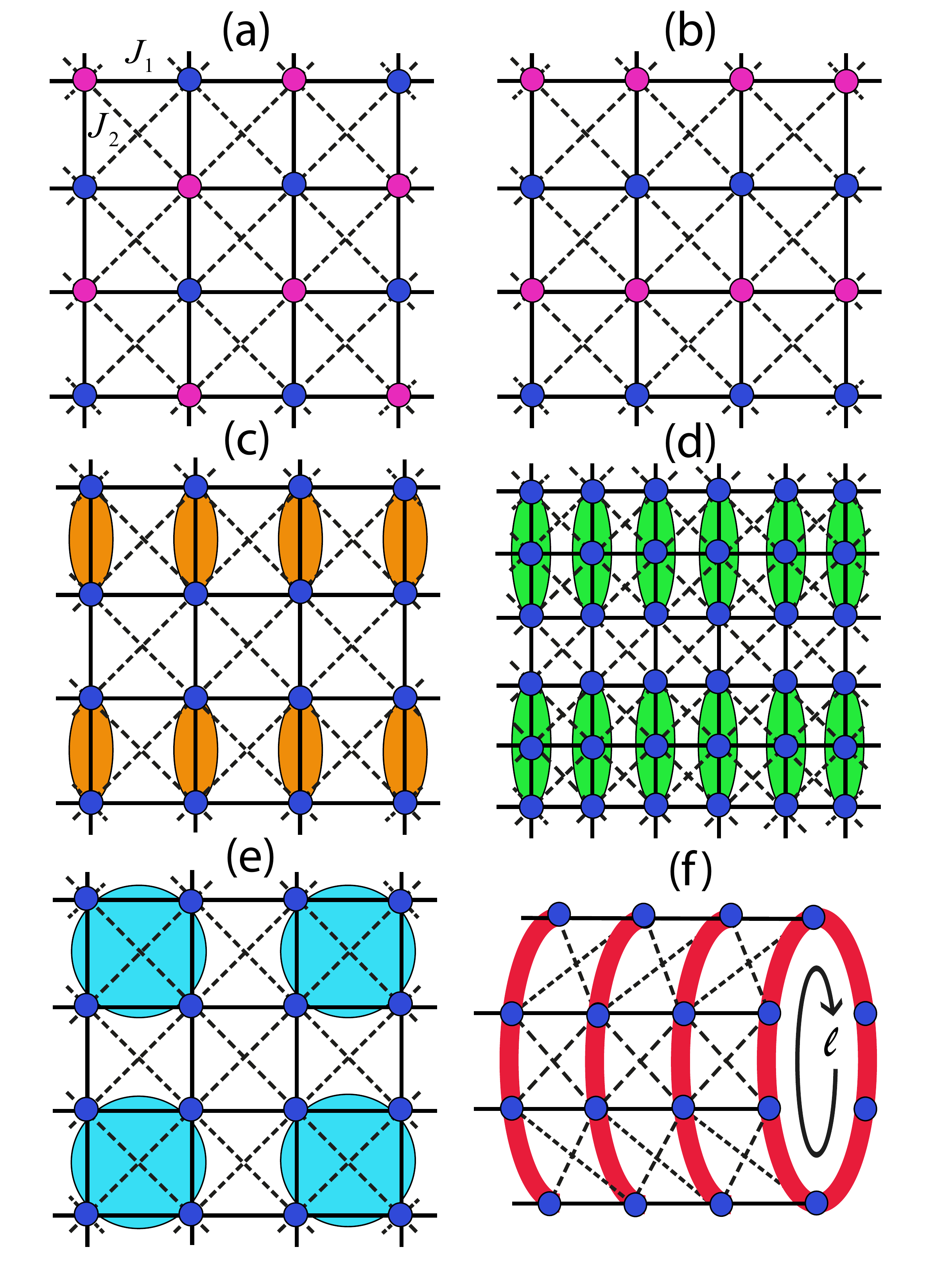}
\caption{(color online) (a),(b) N\'{e}el and striped AFM phases used as 
magnetically 
ordered backgrounds in LSWT. Pink and blue bullets correspond to up and down 
arrangement of spins. (c),(d),(e),(f) Canditates of non-magnetic 
cluster orderings as a ground state background, used in COA. The case of 
(f) corresponds to string-VBS of $\ell=6$. Solid and dashed 
lines are $J_1$ and $J_2$ bonds, respectively. }
\label{fig1}
\end{figure}

In this paper, we therefore examine the spin-$\frac{1}{2}$ transverse field 
Ising model on the $J_1-J_2$ square lattice, Hamiltonian \ref{eq1}, 
by resorting to different analytical and numerical techniques such as linear 
spin-wave theory (LSWT) \cite{Henry:2012}, cluster operator approach (COA) 
\cite{Ganesh:2013,sadrzadeh2015phase} and tree tensor network (TTN) simulation
 \cite{Verstraete-Matrix-2008}.
We found that harmonic quantum fluctuations in LSWT based 
on single-spin flip excitations are incapable of lifting the 
extensive degeneracy of the classical system. However, considering anharmonic 
fluctuations with multi-spin flip excitations via COA
certifies the existence of global-loop-type of quantum fluctuations, 
which are 
able to lift the extensive degeneracy of the system at $J_2/J_1=0.5$ toward a 
string-VBS phase with broken lattice rotational symmetry, leading to an {\it 
order by disorder} transition. 
The string-VBS state is a manifestation of macroscopic quantum 
superposition \cite{Leggett:1980,Abad:2016}.
These findings are further confirmed by 
numerical (TTN) simulations.

The paper is organized as follows. In Sec.~\ref{Model}, we introduce the model 
and some of its classical features. Next, in Sec.~\ref{Methods} we present LSWT 
and COA used for determining the true nature of quantum ground state by 
introducing different type of quantum fluctuations. We compare the results 
obtained from two approaches with each other and also with the TTN results. 
Details of our approaches are presented in Appendices. We argue that 
string-type 
quantum fluctuations can cast the ground state of highly frustrated point 
$J_2/J_1=0.5$ to a string-VBS phase at low fields with broken rotational 
symmetry. 
Sec.~\ref{qpt} discusses the existence of a quantum phase transition from 
string-VBS phase of low fields to a quantum paramagnet phase of high fields at 
$\Gamma/J_1 \cong 0.5$, where the critical exponents are extracted. Finally, 
the paper is summarized and concluded in Sec.~\ref{conclusion}.

\section{Model}
\label{Model}

In this section, we introduce the spin-1/2 transverse field Ising model on the square lattice with $J_1-J_2$ interactions.
We consider a square lattice, where spin-1/2 particles are placed at the 
vertices 
of the lattice and the antiferromagnetic exchange coupling $J_1$ ($J_2$) are 
tuned between the nearest neighbor (next-nearest neighbor) spins (see 
Fig.~\ref{fig1}). Hamiltonian of the model in the presence of a transverse 
magnetic field $\Gamma$ is given by  
\begin{eqnarray}
\mathcal{H}=J_1\displaystyle\sum_{\langle i,j \rangle}{S_i^zS_j^z}
+J_2\displaystyle\sum_{\langle\langle i,j \rangle\rangle}{S_i^zS_j^z}
-\Gamma\displaystyle\sum_{i}{S_i^x},~
\label{eq1}
\end{eqnarray}
where ${S_i}\equiv(S_i^x,S_i^y,S_i^z)$ are the usual quantum spin-1/2 operators 
with ${S_i}^2=S(S+1)$. 

In the extreme case, where $J_2=0$ and $\Gamma=0$, the classical ground state of 
the 
system is given by a N\'{e}el state (Fig.~\ref{fig1}-(a)), which persists as the 
frustration is 
increased up to a critical point at $J_2/J_1=0.5$, where it breaks to a 
collinear anti-ferromagnetic phase with striped AFM order 
(Fig.~\ref{fig1}-(b)) 
for $J_2/J_1>0.5$, 
through a first-order 
quantum phase transition \cite{oitmaa:1981,moran:1993,Kalz:2009}. 
The classical ground state of the system further displays an exponential 
degeneracy at the highly frustrated point $J_2/J_1=0.5$ in which the ground state is 
described by two-up-two-down configurations for spins on every crossed square of 
the lattice. Our aim in this paper is to study the effects of quantum 
fluctuations to lift this extensive degeneracy toward a unique quantum ground 
state. Hence, we consider $J_2/J_1=0.5$ with $\Gamma\neq0$, 
which induces zero-point 
quantum fluctuations to the system due to $S^x$ that does 
not commute with 
other terms in the Hamiltonian \ref{eq1}.

\section{Nature of quantum fluctuations}\label{Methods}
\subsection{Linear Spin Wave Theory}
\label{LSWT}

To incorporate harmonic quantum fluctuations within LSWT, we start with 
the degenerate classical magnetically ordered backgrounds at 
$J_2/J_1=0.5$, i.e. N\'{e}el and striped AFM 
phases shown in Fig.~\ref{fig1}-(a, b). The transverse magnetic field, 
$\Gamma$, creates the same canting angle $\theta$ on each classical spin vector 
of the N\'{e}el and striped configurations. Accordingly, the classical spin 
components become $S^{x}_{i}=S\sin\theta$ and $S^{z}_{i}=\pm S\cos\theta$, 
where 
$\pm$ sign denotes up and down spins in the $z$-direction. The angle 
$\theta$ increases with the strength of the transverse field $\Gamma$ up to the 
maximum value of $\theta_{max}=\pi/2$ , which corresponds to a full 
polarization of the classical spins in the $+x$-direction for 
$\Gamma\geq\Gamma_c$, where $\Gamma_c$ is the critical magnetic field. 
For a system with $N$ spins, there are $2N$ bonds with $J_1$ coupling 
and 2N bonds with $J_2$ coupling on the square lattice. The classical ground state energy per spin for both the N\'{e}el and 
striped phases are therefore given as
\begin{eqnarray}
\varepsilon_{cl}^{N\acute{e}el}=\varepsilon_{cl}^{striped}
=-S^2cos^2\theta-\Gamma S sin\theta. 
\label{eq2}
\end{eqnarray}
After minimizing the classical energy per spin with respect to angle $\theta$, 
we set $\theta=\pi/2$ to obtain the critical transverse field 
$\Gamma_c^{LSWT}$, given by $\Gamma_c^{LSWT}=2S$. Then, a LSWT is constructed on 
each of the 
two classical canted N\'{e}el and canted striped AFM magnetically ordered backgrounds. 
Harmonic quantum fluctuations of LSWT around these classical reference states 
will reduce the magnitude of the classical order parameters and result in 
zero-point energy corrections.
In a general formalism (see Appendix.~\ref{ap-lswt}), 
we define $S_{l,p}$ as the {\it p}-th spin ($p=1,\ldots,n$) of the {\it l}-th 
cell, where $n$ is the number of spins in a magnetic unit cell. We consider 
small quantum fluctuations on the classical reference states by linearized 
Holstein-Primakoff transformations and finally obtain an effective diagonal 
quadratic form of Hamiltonian Eq.~\ref{eq1} as,
\begin{equation}
\label{eq3}
\mathcal{H}_{LSWT} = 
E_{cl}-\dfrac{N}{n}\displaystyle\sum_{p}{\left(\frac{\tilde{h}_p}{2}\right)}
+\displaystyle\sum_{\textbf{k},p}{\omega_{\textbf{k},p}\left(c_{\textbf{k},p}^{
\dagger}c_{\textbf{k},p}+\dfrac{1}{2}\right)},
\end{equation}
where $\textbf{k}$ sums over the first Brillouin zone of the lattice constructed 
from the centers of magnetic unit cells of the classical reference state. Furthermore, $p$ 
runs over the n spins of a magnetic unit cell, $\tilde{h}_p$ is a quantum correction
and $\omega_{\textbf{k},p}$ 
define the spectrum of quasi-particles, which are created by the 
bosonic creation 
operators $c_{\textbf{k},p}^{\dagger}$. 
The effective Hamiltonian, Eq.~\ref{eq3}, obtained within LSWT framework, shows 
that both N\'{e}el and striped magnetically ordered backgrounds have the same 
zero-point energy corrections, which is a result of the harmonic 
single-spin-flip excitations. Hence, quantum corrections at harmonic 
level do not distinguish between different ordered manifold of states, failing 
to lift the extensive degeneracy at $J_2/J_1=0.5$. Moreover, as shown 
in 
Fig.~\ref{fig5}, we observe a violation of Hellmann-Feynman theorem at enough high fields 
before reaching the critical point $\Gamma_c^{LSWT}/J_1=1.0$. Indeed, by 
increasing the transverse field $\Gamma$, before reaching
the critical value, the transverse magnetization obtained from 
Hellmann-Feynman theorem, $m_x=-\frac{1}{S}\frac{\partial\langle 
H\rangle}{\partial \Gamma}$, deviates from the expectation value of 
magnetic order parameter $m_x=\frac{1}{S}\langle S_x\rangle$, signaling a
violation of the Hellmann-Feynman theorem. This
inconsistency implies again that quantum fluctuations go beyond the harmonic level of approximation considered in LSWT.

\subsection{Cluster Operator Approach}
\label{COA}

In order to consider anharmonic quantum fluctuations, we implement the cluster operator 
approach. Analogous to the spin-wave theory, a candidate cluster-ordered background is proposed 
above which, the anharmonic multi-spin excitations are defined. This is in contrast to the LSWT, 
where only single-spin excitations have been taken into account. 
Let us further note that COA besides the introduction of anharmonic quantum 
fluctuations, can reveal the existence of possible valence bond solid 
phases. Generally, VBS phases are appeared as a regular 
pattern of dimers, trimers, quadrumers or loops shown in Fig.~\ref{fig1}-(c-f). 
 
It is shown that at 
zero field and $J_2/J_1=0.5$, two-spin flip
excitation on a dimer, three-spin flip excitation on a trimer or four-spin flip excitation on a quadrumer 
cost the same finite energy as a single-spin flip one, i.e. $4J_1$. This is 
true for any finite cluster, which is shown in Appendix.~\ref{ap-eszf}. 
However, 
flipping the spins on a vertical or horizontal 
global closed loop 
costs zero excitation energy, keeping the system in the degenerate ground 
state manifold (see Appendix.~\ref{ap-eszf}). Therefore, it can be anticipated
that in the presence of quantum fluctuations by a transverse field $\Gamma$, such global 
loops are proper building blocks to construct the ground state structure 
of the model. To confirm such assertion, we consider four 
candidate cluster orderings shown in Fig.~\ref{fig1}-(c-f) as ground state 
backgrounds used in COA. We obtain an effective theory by a bosonization 
formalism for each cluster configuration, and then compare their results 
with each other to confirm that the true excitations of the model are of the 
global-loop type, constructing a columnar string-VBS phase for low fields at 
the highly frustrated point.

The following steps are carried out to construct an effective theory for the 
candidate cluster ordered backgrounds of Fig.~\ref{fig1}. First, we rewrite the 
Hamiltonian, Eq.~\ref{eq1}, as a sum over two terms,
$\mathcal{H}=\mathcal{H}_0+\mathcal{H}_{int}$, where 
$\mathcal{H}_0=\sum_I{\mathcal{H}}_{I}$ denotes the set of shaded isolated 
clusters and $\mathcal{H}_{int}$ defines the interaction Hamiltonian between 
them. Next, we associate a boson to each eigenstate of the TFI Hamiltonian on a 
single cluster. In this respect, each eigenstate $\vert u\rangle$ of cluster I 
is created by a boson creation operator $b_{I,u}^\dagger$ acting on the vaccum 
$\vert0\rangle$, i.e. $\vert u\rangle_I=b_{I,u}^\dagger\vert0\rangle$, where 
$b_{I,u}^\dagger$ and $b_{I,u}$ are usual bosonic operators, satisfying 
$[b_{I,u},b_{I,u}^\dagger]=1$ and $[b_{I,u}^{(\dagger)},b_{I,u}^{(\dagger)}]=0$. 
Hence, a cluster ordered background is a 
Bose-condensate of ground state $\vert u=1\rangle$ bosons, i.e. 
\begin{eqnarray}
\langle b_{I,1}^\dagger b_{I,1} \rangle \equiv\bar{p}^2,\quad \forall I
\label{eq4}
\end{eqnarray}
where $\bar{p}$ is the condensation amplitude and $\bar{p}^2$ gives the 
probability of such condensation. In the absence of inter-cluster interactions, 
$\bar{p}^2$ is equal to unity. Therefore, the Bose-condesate background acts 
like an ordered-reference state, above which quantum 
fluctuations will reduce the magnitude of condensation probability $\bar{p}^2$ 
and result in zero-point energy corrections. In other words, inter-cluster 
interactions give rise to low-lying excitations above the perfect cluster 
ordered background. As a result of hybridization of ground state of each 
cluster to other excited states, the value of $\bar{p}^2$ reduces from 
unity by bringing about a non-zero occupation of other excited bosons. 
Nevertheless, for preserving the Hilbert space of the effective model, the total 
occupation of bosons per cluster should be equal to one. According to these 
arguments, the effective Hamiltonian for a cluster ordered background can now be 
written in a quadratic bosonic form within a mean-field approximation of 
condensated bosons, as
\begin{eqnarray}
\nonumber \mathcal{H}&=&N_c\bar{p}^2\epsilon_{1}+\sum_{I}\sum_{u}\epsilon_ub_{I,u}^\dagger b_{I,u}\\
 &&-\mu\Big[N_c\bar{p}^2 +\sum_{I,u\neq1}b_{I,u}^\dagger b_{I,u}-N_c\Big]\\
\nonumber &&+\bar{p}^2\sum_{\langle I,J\rangle}\sum_{u,v} 
\Big[(d_{uv}^{IJ})\:b_{I,u}^\dagger b_{J,v}^\dagger+(h_{uv}^{IJ})\: 
b_{I,u}^\dagger b_{J,v}+ h.c.\Big].
\label{eq5}
\end{eqnarray}
The first line includes intra-cluster terms, where the index $I$ sums over all isolated clusters, $u$ sums over the dominant excited states of each cluster with corresponding eigen energies $\epsilon_u$, and $N_c$ denotes the total number of isolated clusters. The second line enforces single boson occupancy constraint, via a chemical potential $\mu$. The third line involves inter-cluster terms, where $u,v$ are the two excited bosons of neighboring clusters $I$ and $J$, respectively. Coefficients $d_{uv}^{IJ}=\langle uv\vert\mathcal{H}_{IJ}\vert11\rangle$ 
and $h_{uv}^{IJ}=\langle u1\vert\mathcal{H}_{IJ}\vert1v\rangle$ are respectively creation and 
hopping amplitudes between excited bosons of neiboring clusters.
The minimization of ground-state energy of the bosonic effective theory in 
addition to the conservation of Hilbert space dimension are satisfied by
the solution of $\frac{\partial \langle \mathcal{H} \rangle}{\partial \mu} = 0$ and $\frac{\partial \langle \mathcal{H} \rangle}{\partial \bar{p}} = 0$ equations. Details of the bosonic 
effective theory for COA are given in Appendix.~\ref{ap-coa}.

\begin{figure}
\includegraphics[width=\columnwidth]{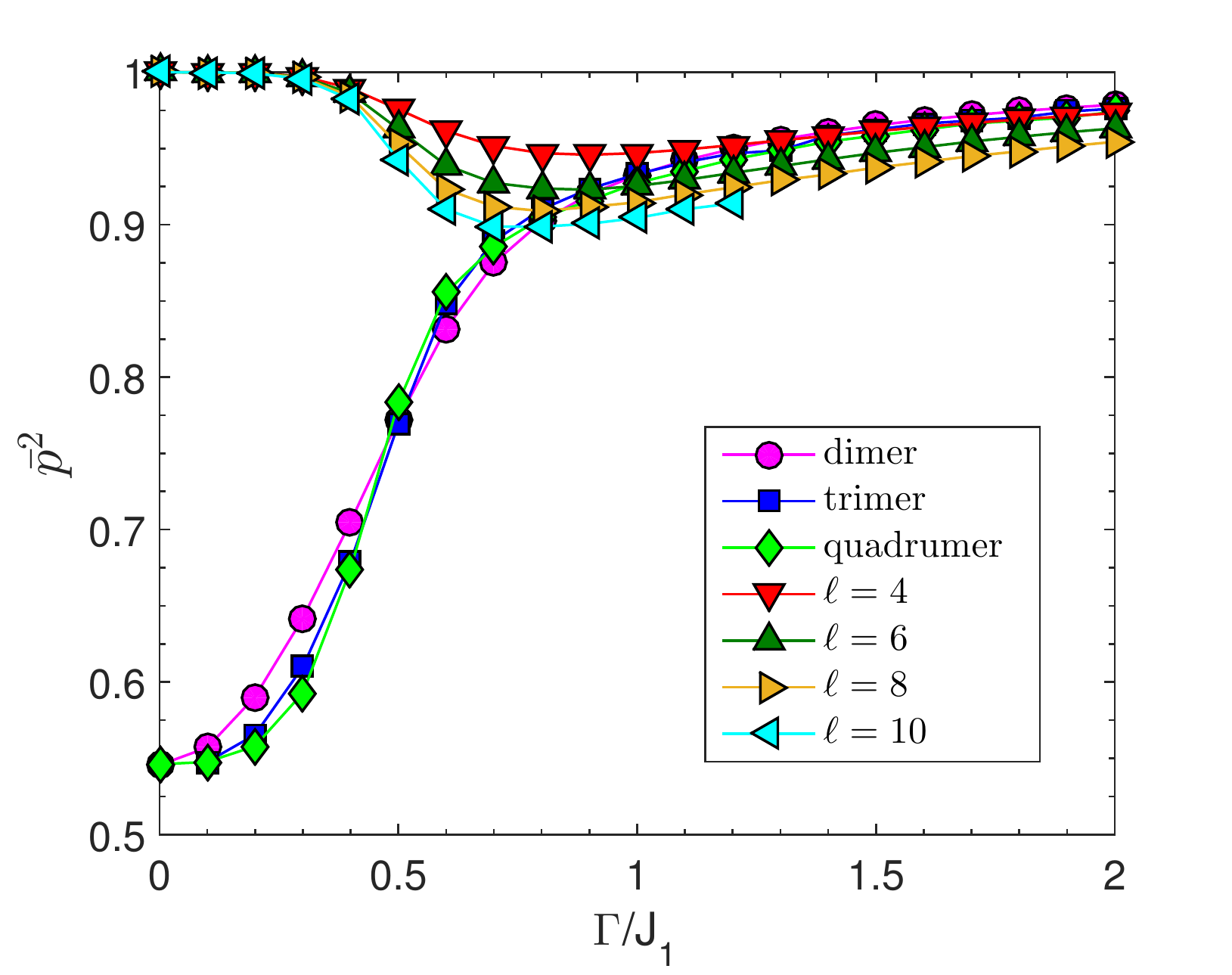}
\caption{(color online) The condensation probability $\bar{p}^2$ for different 
cluster orderedings of Fig.~\ref{fig1}-(c, d, e, f). $\ell$ denotes the 
cylinder perimeter length considered in COA for a string-ordered background 
shown 
in Fig.~\ref{fig1}-(f)}
\label{fig2}
\end{figure}

The condensation probability $\bar{p}^2$ of different cluster orderings shown in Fig.~\ref{fig1}-(c, d, e, f) is demonstrated in Fig.~\ref{fig2}. We found that for low transverse fields
($\Gamma <0.5$), there is a strong condensation 
probability (near unity) of the global loops 
($\ell=4, 6, 8, 10$) on the lattice, while
the condensation probability of dimers, trimers and 
plaquettes (quadrumers) is weak ($\sim 0.55$). 
We have also considered the staggered dimer configuration in our
calculations not shown here, which gives a result similar to the dimer case.
Let us note that a global loop is a closed string,
which covers all sites along a horizontal or vertical direction of 
the periodic square lattice, as shown
in Fig.~\ref{fig1}-(f).
This implies that at low fields
the proper conjecture for the ground-state structure is based on the 
global loops, while finite size clusters fail to condensate properly 
in the ground state.
%However, when we go to the high-field regimes ($\Gamma >1$)
%any kind of clusters can be used to explain a quantum paramagnet phase
%in terms of COA, as they all condense very well to the ground 
%state as shown in Fig.~\ref{fig2}.

\begin{figure}
\includegraphics[width=\columnwidth]{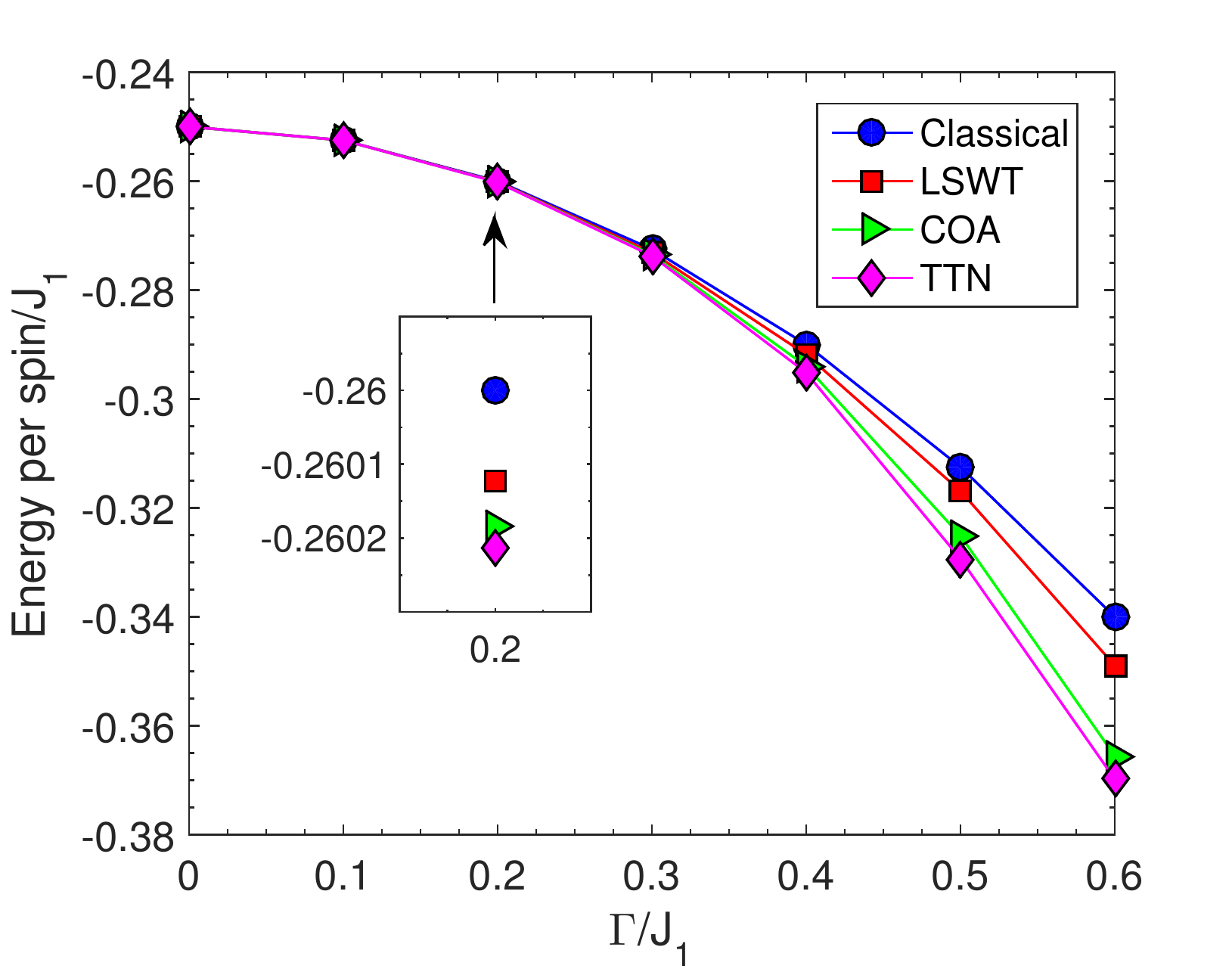}
\caption{(color online) Ground-state energy per site versus transverse magnetic field.
A comparision between classical, linear-spin wave, cluster operator (with 
string-ordered background on $\ell=10$ cylinder) and 
tree-tensor network (on a 8 $\times$ 6 lattice) approaches is presented. The inset shows the magnified data for $\Gamma/J_1=0.2$.}
\label{fig3}
\end{figure}

The results of Fig.~\ref{fig2} suggest a string-VBS phase for the ground state 
of our model at low fields. It turns out that anharmonic quantum fluctuations, 
mediated in terms of COA, lift the classical degeneracy at 
the highly frustrated point, towards a string-VBS phase, which breaks lattice 
rotational symmetry and leaves the system with a two-fold degenerate ground state. 
The ground-state energy per site versus transverse magnetic field is illustrated
in Fig.~\ref{fig3}. We observe that the energy of the string-VBS state obtained 
from 
COA is less 
than the classical and LSWT ones, justifying the existence of string formation
in the ground state. The inset of Fig.~\ref{fig3} clearly shows the lower energy 
value of COA for a low field value $\Gamma/J_1=0.2$. We have also shown the 
ground-state energy obtained from TTN numerical algorithm, as a 
reference close 
to the exact diagonalization data. The numerical TTN is a renormalization 
ansatz to simulate large lattice sizes that is explained in Appendix.~\ref{ap-ttn}. 
Accuracy of our data on $\{4\times 6, \, 6\times 6, \, 6\times 8 \}$ lattices is of order, respectively, $\{10^{-8}, \, 10^{-5}, \, 10^{-4} \}$ which is not presented here.
\begin{figure}
\includegraphics[width=\columnwidth]{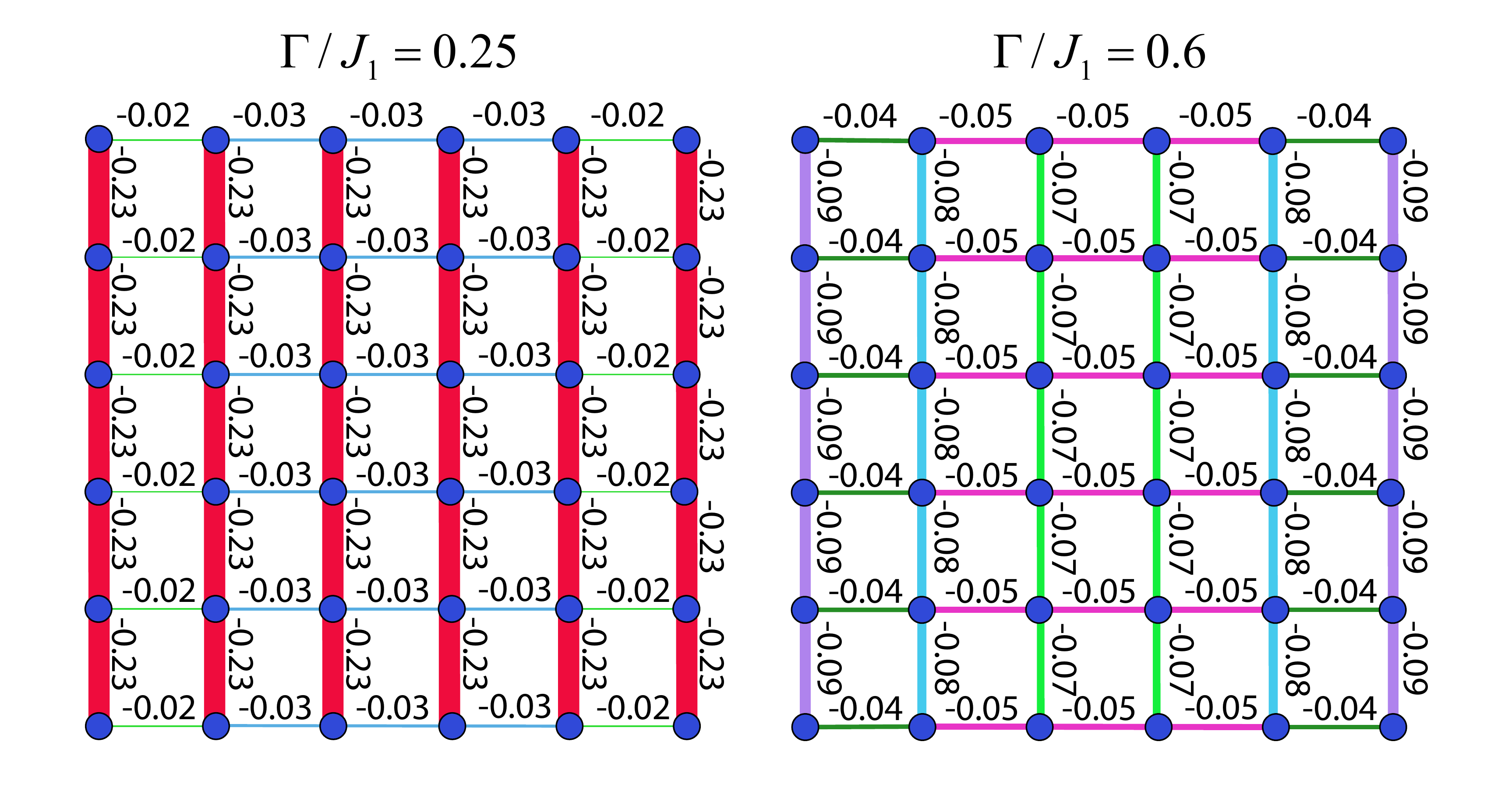}
\caption{(color online) Nearest-neighbor correlations, $C_{NN}$, obtaind by TTN 
numerical simulation, which measures the breaking of lattice rotational 
symmetry in the string-VBS phase. Left panel: low-field regime representing the 
string-VBS state. Right panel: high-field regime of the quantum paramagentic 
phase, which preserves rotational symmetry.}
\label{fig4}
\end{figure}

The nature of ground state can be represented by the nearest-neighbor 
(NN) correlation
function,
\begin{equation}
C_{NN}=\langle S_i^z S_j^z \rangle, \;\;\; i, j: \mbox{NN on the lattice}.
\label{eq6}
\end{equation}
In this respect, we compute $C_{NN}$ on a 6$\times$6 lattice using TTN 
which is shown in Fig.~\ref{fig4}, for two different 
values of transverse field $\Gamma$. The left panel of Fig.~\ref{fig4}
corresponds to low-field regime ($\Gamma/J_1=0.25$), while the right panel
corresponds to the high-field values ($\Gamma/J_1=0.60$). The left panel shows that the correlations along the vertical 
direction are close to their maximum value of N\'{e}el type ordering
($|C_{NN}^{max}|=0.25$), while
the correlations on the horizontal direction is very small. This is 
a clear signature of the string formation as a VBS phase. The 
emergence of strings could be either in vertical or horizontal direction, 
breaking 
the rotational symmetry of the lattice which manifests the two-fold degeneracy. 
Increasing the magnetic field to the high-field regime drives the model
to a quantum paramagnet, which has rotational symmetry and leads to almost
equal correlations along the two perpendicular directions as shown in the right 
panel of Fig.~\ref{fig4}. Such symmetry breaking of ground state at low fields is
a signature for presence of a quantum phase transition from the string-VBS 
phase of low fields to the
quantum paramagentic phase of high fields, which is investigated with more details in the next section.

\section{Quantum phase transition}\label{qpt}

\begin{figure}
\includegraphics[width=\columnwidth]{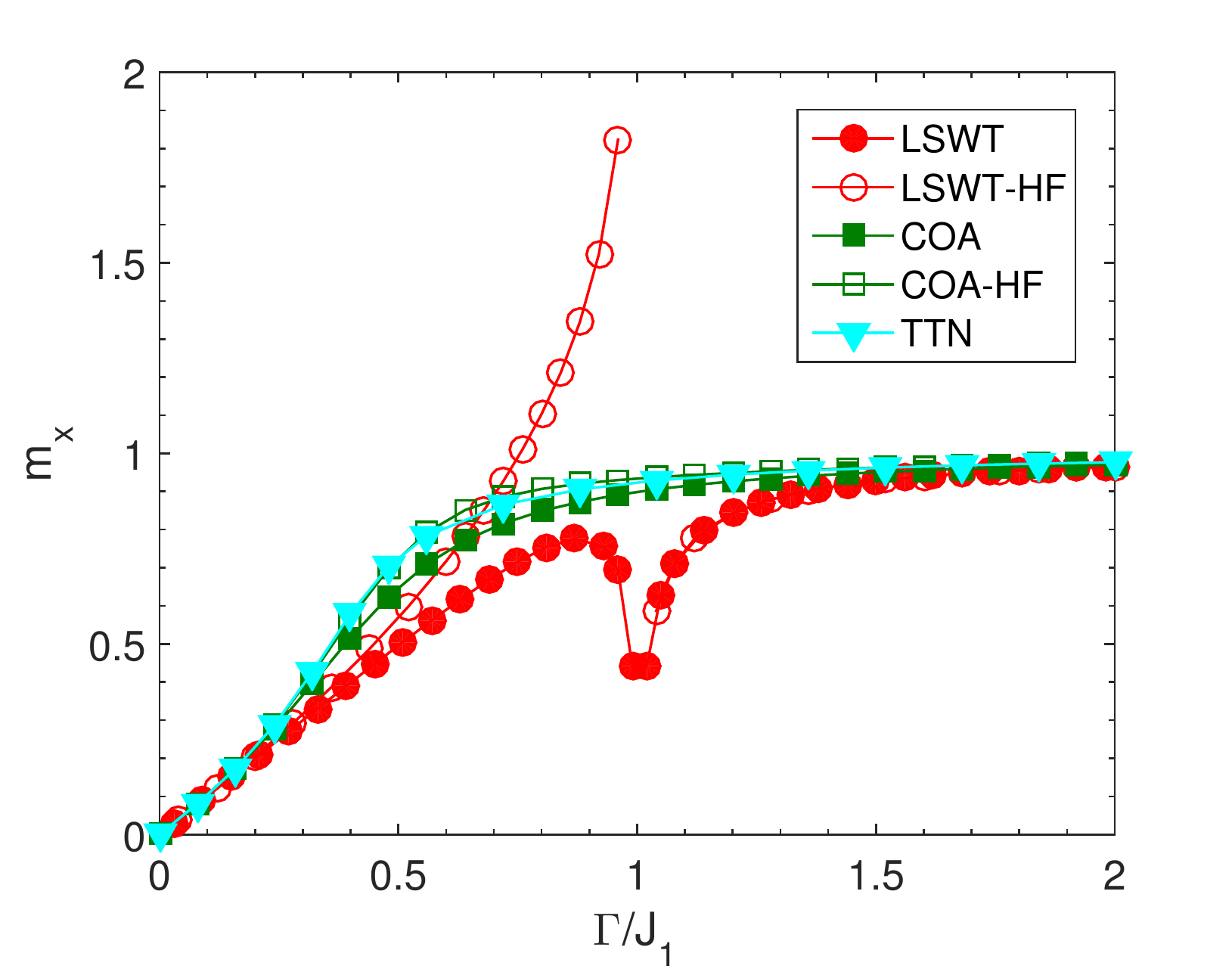}
\caption{(color online) Transverse magnetization calculated by different 
methods, LSWT, COA and TTN. 
The value obtained directly from the 
expectation value of $S_x$ operator, $m_x=\frac{1}{S}\langle S^x\rangle$ is 
compared with the one obtained from Hellmann-Feynmann theorem, 
$m_x^{HF}=-\frac{1}{S}\frac{\partial\langle H\rangle}{\partial \Gamma}$.}
\label{fig5}
\end{figure}

In this section, we study the behavior of field induced magnetization, $m_x=\frac{1}{S}\langle S^x\rangle$, by increasing the transverse
field from the low-field to high-field regimes. The magnetization as a function 
of transverse field calculated by different approaches is depicted in 
Fig.~\ref{fig5}. Moreover, the transverse magnetization $m_x$, obtained directly 
from the expectation value of $S_x$ operator, is compared with the derivative 
of 
effective Hamiltonian expectation value versus $\Gamma$, i.e. 
$m_x^{HF}=-\frac{1}{S}\partial\langle 
H\rangle/\partial \Gamma$ corresponding to Hellmann-Feynmann theorem. As we mentioned in Sec.~\ref{LSWT}, LSWT 
exhibits a violation of the Hellmann-Feynmann theorem as the magnetic field increases toward the high-field regime. This implies that when increasing the transverse field
Γ, quantum fluctuations are beyond the harmonic level
of approximation considered in LSWT. However, COA results are in a good 
agreement
with the Hellmann-Feynmann theorem and numerical results obtained from TTN simulation. 
The COA results further show that the anharmonic quantum 
fluctuations will render the violated region of the Hellmann-Feynmann theorem to 
the quantum paramagnet phase, proposing a lower critical value than the LSWT 
counterpart, between the string-VBS and quantum paramagnet phases. 

\begin{figure}
\centering
\includegraphics[width=\columnwidth]{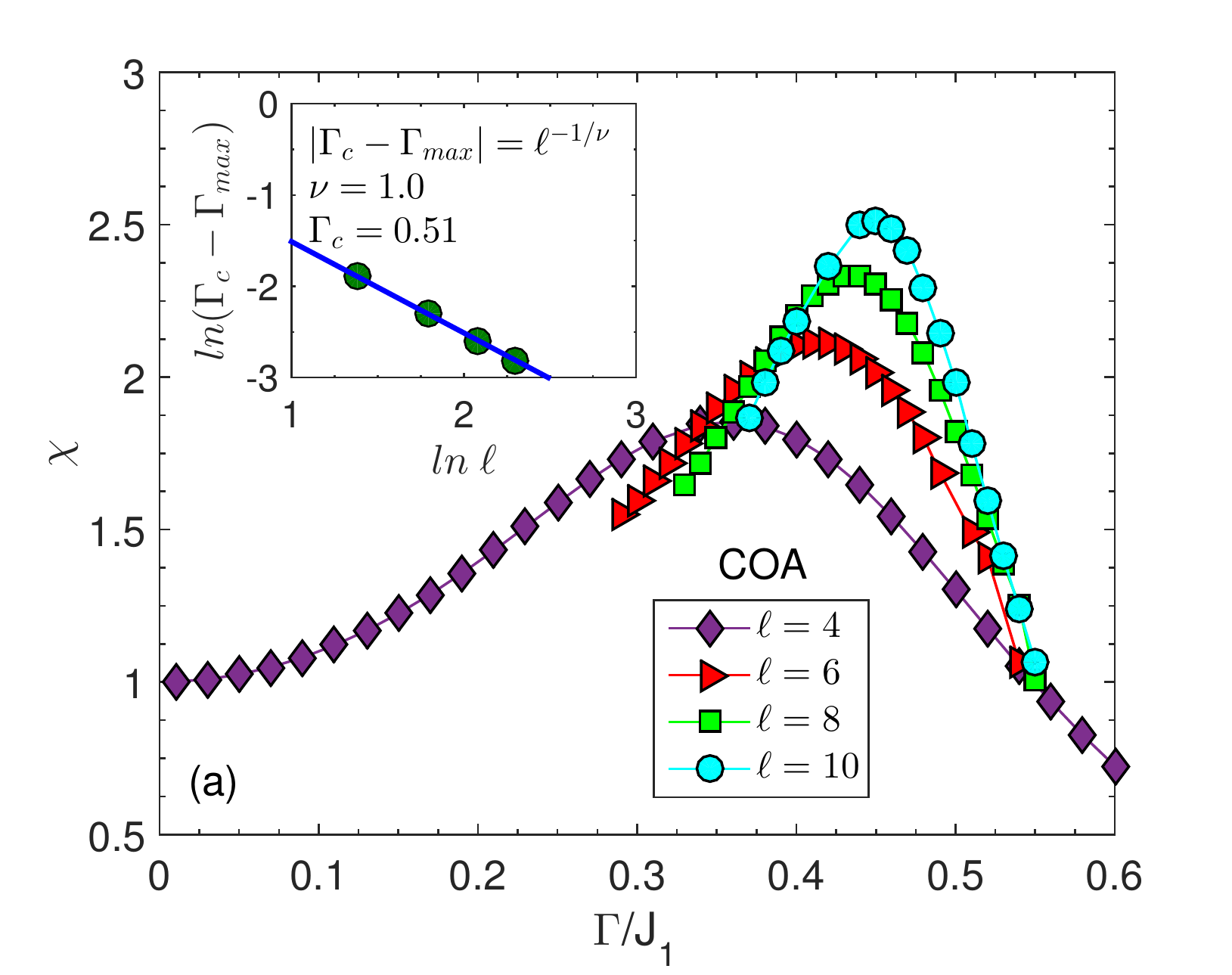}
\includegraphics[width=\columnwidth]{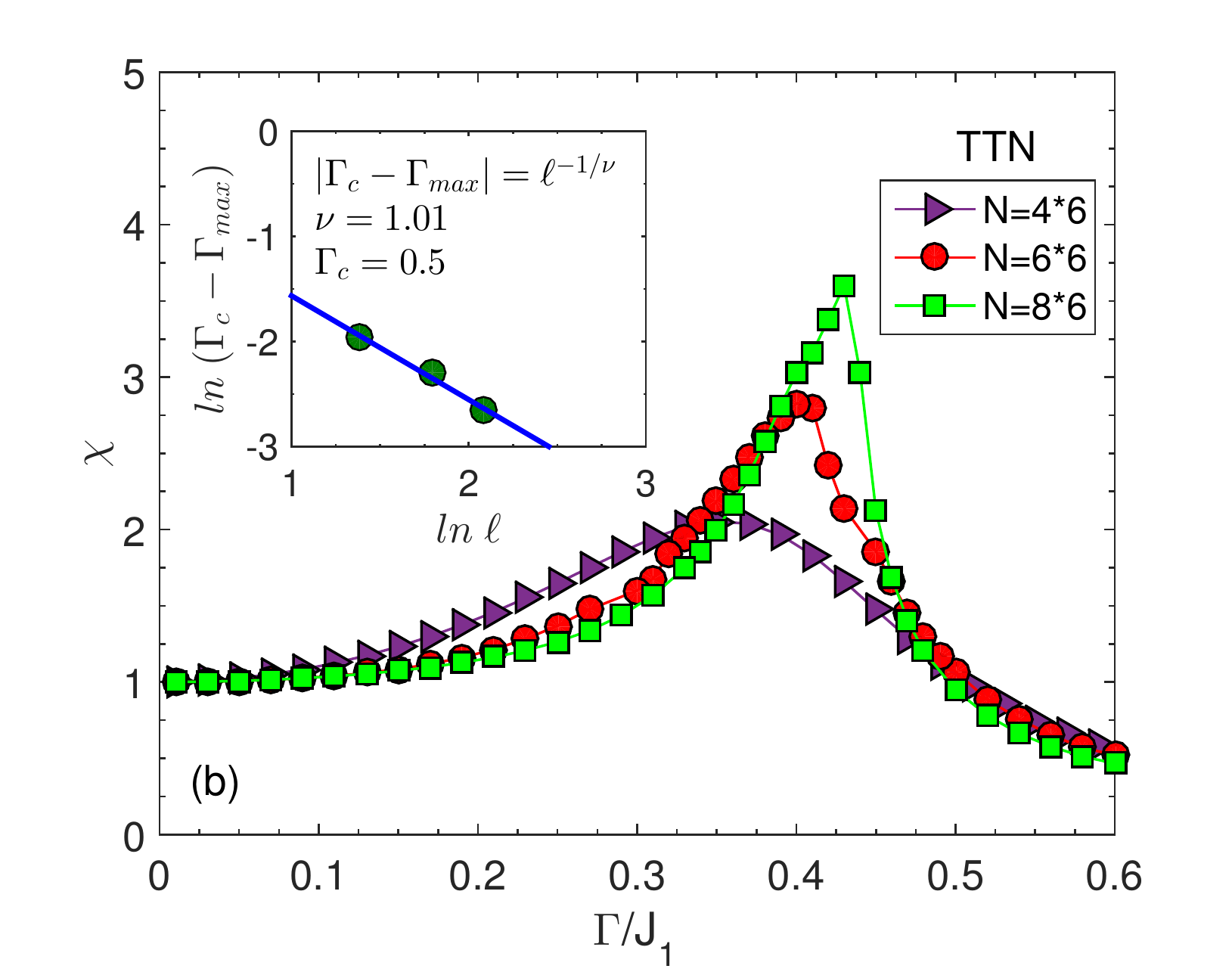}
\includegraphics[width=\columnwidth]{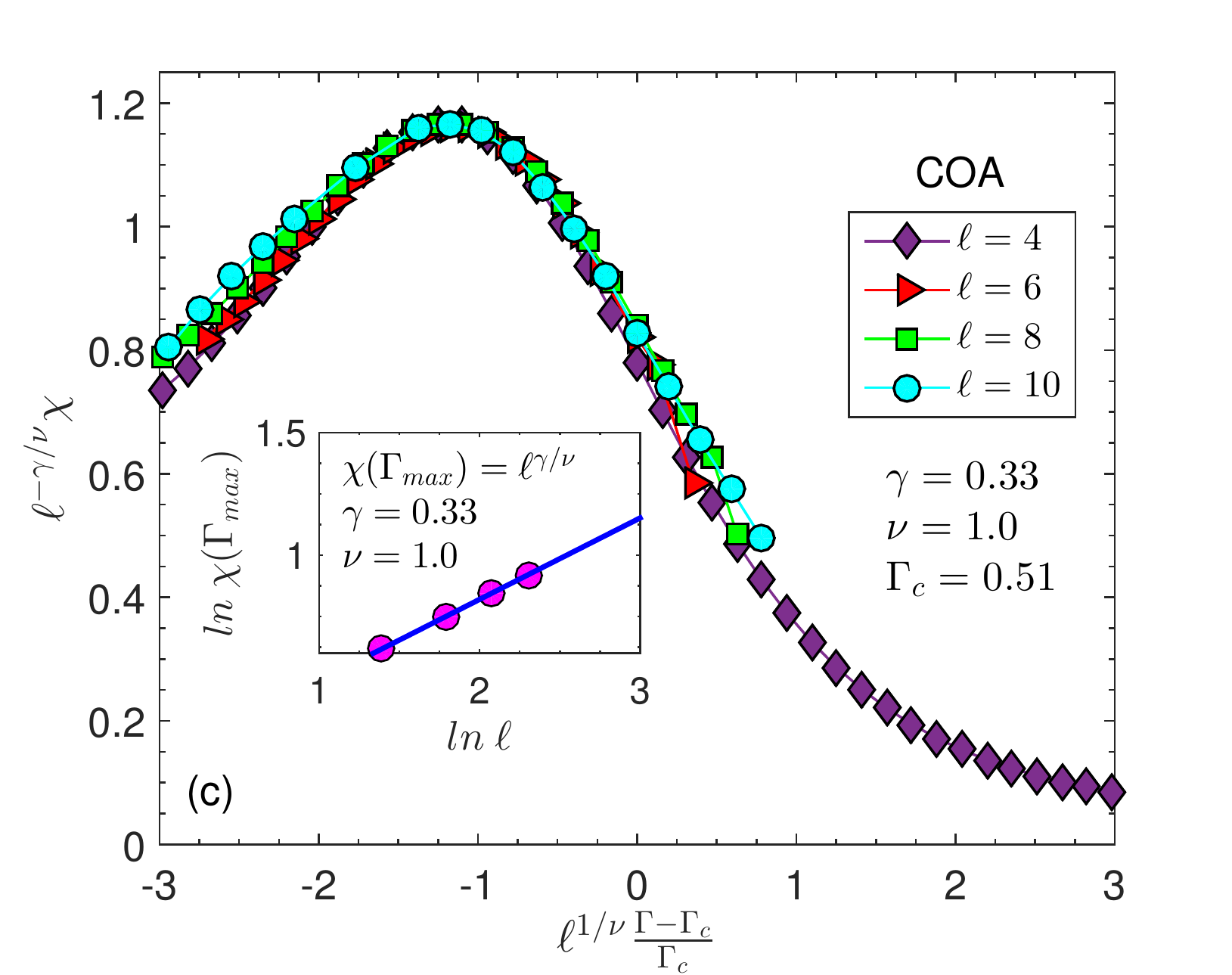}
\caption{(color online) (a), (b) 
Magnetic susceptibility
obtained from COA and TTN, 
respectively for different system sizes. They show a sharp peak indicating a 
phase transition from string-VBS 
phase of low fields to the quntum paramagnet of high fields, 
at $({\Gamma/J_1})_c = 0.5\pm 0.01$ 
with exponent $\nu=1.0 \pm 0.01$. (c) Data 
collapse of magnetic 
susceptibility obtained from COA, which shows the scale invariance of 
susceptibility governed by exponent $\gamma=0.33\pm 0.01$. }
\label{fig6}
\end{figure}

Quantum phase transition can be traced out by the divergent behavior 
of magnetic susceptibility, which is the derivative of transverse magnetization 
with respect to the magnetic field, 
\begin{equation}
 \chi = \frac{\partial m_x}{\partial \Gamma}=-\frac{1}{S}\frac{\partial^2E}{\partial\Gamma^2}.
 \label{eq7}
\end{equation}
The magnetic susceptibility of COA data is plotted in Fig.~\ref{fig6}-(a),
which shows sharper and stronger divergence as the length ($\ell$) of lattice 
is 
increased. Let us note that the COA results with string-ordered background are 
obtained for lattices defined on an infinite 
cylinder that has a finite perimeter length $\ell$. Finite-size scaling theory 
tells us how to estimate the critical exponents
for the model \cite{Nishimori:2011}. The divergent behavior of $\chi$ obeys the following 
scaling relations  
\begin{eqnarray}
 \label{eq8}|\Gamma_c - \Gamma_{max}| &\sim & \ell^{-1/\nu},\\
 \chi(\Gamma_{max})&\sim & \ell^{\gamma/\nu},
 \label{eq9}
\end{eqnarray}
where $\Gamma_c$ is the critical field in the thermodynamic limit,
$\Gamma_{max}$ is the position of maximum of finite-lattice susceptibility 
$\chi$, $\nu$ is the correlation length exponent
i.e. $\xi \sim |\Gamma - \Gamma_c|^{-\nu}$ and $\gamma$ is an exponent, which
governs singularity in the magnetic susceptibility. As shown in the inset of 
Fig.~\ref{fig6}-(a), we find a good scaling for COA data, which gives $\nu=1.0 
\pm 0.01$ and 
$\Gamma_c=0.51 \pm 0.01$. A similar behavior is also observed for $\chi$ versus $\Gamma$ of 
the TTN numerical computation presented in Fig.~\ref{fig6}-(b). 
Accordingly, the same critical field and exponent $\nu$ are also reported in 
the inset
of Fig.~\ref{fig6}-(b). Once we have obtained $\nu$ and $\Gamma_c$
from Eq.~\ref{eq8}, we can use them to find $\gamma$ from Eq.~\ref{eq9},
as well as getting the scale invariant behavior of magnetic susceptibility,
which is observed from a good data collapse of different sizes in 
Fig.~\ref{fig6}-(c). It shows the scale invariance of susceptibility
with exponent $\gamma=0.33  \pm 0.01$. 
Both COA and TTN imply a continuous phase transition from 
string-VBS phase (at low fields) to the quantum paramagnet phase (at high 
fields) at $\Gamma_c=0.5\pm0.01$. The continuous nature of such transition is confirmed by the broken lattice 
rotational symmetry in the string-VBS phase compared with symmetric quantum 
paramagnet phase. 

We would like to comment on the nature of string-VBS ground state.
The formation of loops yields the ground state to inherit partially the 
one-dimensional (1D) character of TFI model.
At zero field, the ground state of 1D TFI model is doubly degenerate, which
is given by classical antiferromagnetic state $|\phi\rangle=|+-+- \dots 
+-\rangle$ and its spin flipped one
$|\bar{\phi}\rangle=|-+-+ \dots -+\rangle$, where $|+\rangle, |-\rangle$ 
represent the eigenstates of $S^z$ Pauli operator. 
In the presence of small transverse field, the ground state is a linear 
superposition of different configurations mostly occupied by  $|\phi\rangle$
and $|\bar{\phi}\rangle$. This is actually a macroscopic superposition
of quantum states, which has been discussed by Leggett \cite{Leggett:1980}
to distinguish between macroscopic quantum superposition and quantum 
entangelement. A recent study in Ref.~\cite{Abad:2016} verifies that the
ground state of 1D TFI model in AFM region is essentially a superposition
of the two macroscopic distinct states $|\phi\rangle$ and $|\bar{\phi}\rangle$,
i.e. a macroscopic quantum superposition. We therefore conjuncture that the 
string-VBS state
is a witness for two-dimensional version of macroscopic quantum superposition. 
In other words, we conclude that string-VBS phase consists of a columnar 
ordering of 
string-valence bonds each of which in an equal superposition of two possible 
N\'{e}el configurations with no magnetic order in z-direction.

\section{Summary and Conclusions}
\label{conclusion}
We have studied the zero-temperature phase diagram of the transverse field Ising 
model on the $J_1-J_2$ square lattice at the highly frustrated point 
$J_2/J_1=0.5$, which is known to have an extensive degenerate classical ground 
state at $\Gamma=0$. 
The LSWT analysis of the model failed to lift 
this classical degeneracy implying that harmonic fluctuations, coming 
from the single-spin flip excitations, are not able to represent the true 
quantum fluctuations 
of the system at the highly frustrated region. We therefore, 
applied the cluster operator 
approach, which is based on the multi-spin flip type of anharmonic 
quantum fluctuations above a non-magnetic cluster ordered background. We found 
that the exponential degeneracy of the 
classical ground state at $J_2/J_1=0.5$ is lifted toward a string-VBS phase 
which 
breaks rotational symmetry of the lattice with only two-fold degeneracy.  
This is a manifestation of order-by-disorder transition that is induced by 
anharmonic quantum fluctuations. 

The quantum phase transition between string-VBS 
phase at low fields and quantum paramagnet phase at high fields occurs at the
critical point $(\Gamma/J_1)_c=0.50\pm0.01$ and is of a continuous type 
as the rotational symmetry is only broken at the string-VBS phase. 
The critical exponents have been obtained to be 
$\nu=1.0\pm0.01$ and $\gamma=0.33\pm0.01$.
Moreover, we conjuncture that the string-VBS state is an example of macroscopic 
superposition
of distinct quantum states in 2D, where 
the whole lattice is a direct product of 1D ground states, i.e.
$\bigotimes_j \big(|\phi_j\rangle + |\bar{\phi}_j\rangle \big)$.

Let us discuss the connection of our results to the phase diagram 
of spin-1/2 $J_1-J_2$ AFM Heisenberg model on two-dimensional square lattice.
The ground state structure of Heisenberg model at $J_2/J_1=0.5$
is controversial to be either a valence bond solid state or a spin liquid
phase 
\cite{singh:1999,metavitsiadis:2014,Isaev:2009,yu:2012,doretto2014plaquette,
hu:2013,wang:2013,Gong:2014PRL,morita:2015,jiang:2012,mezzacapo:2012,li:2012,
ren2014cluster}. Early studies proposed that anharmonic fluctuations could 
make a dimer-VBS \cite{PhysRevB.44.12050} or a plaquatte-VBS 
\cite{PhysRevB.54.9007} as stable phases around $J_2/J_1=0.5$, 
granting that short-range corrections to the ground-state energy 
are small. Our COA results on TFI model with dimer-VBS is similar
to the case of Ref.\cite{PhysRevB.44.12050}, where dimer-VBS corrections
are not small to construct a stabilized dimer-VBS at $J_2/J_1=0.5$.
According to the recent investigations, a quntum spin liquid is more plausible 
phase for the intermediate 
region of the Heisenberg model
\cite{hu:2013,wang:2013,Gong:2014PRL,morita:2015,jiang:2012,mezzacapo:2012,
li:2012,
ren2014cluster}.
On the other hand, our results on TFI model govern the high 
anisotropy limit of the Heisenberg model, where 
the easy-axis coupling is much stronger than the coupling in 
the fluctuating plane. It suggests that we get the string-VBS ground state
by increasing the easy-axis anisotropy of the Heisenberg model.
In other words, we conclude that by reduction of symmetry 
from $SU(2)$ to $Z_2$, plausible spin liquid phase of $J_1-J_2$ Heisenberg 
model on the square lattice cast to a 
string-VBS phase at the highly frustrated point $J_2/J_1=0.5$. Such a novel 
string-VBS phase can also emerge in the case of 
reducing quantum fluctuations by increasing the dimensionality or the spin 
quantum number, as it was predicted in previous literature for a $S=1$ 
$J_1-J_2$ 
Heisenberg model on the square lattice \cite{cai2007two,jiang2009phase}.

\section{Acknowledgements}
S.S.J. and A.L. acknowledge support from the Iran National Science Foundation 
under Grant No. 93023859 
and Sharif University of Technology’s Office of Vice President
for Research.

\appendix
\section{Linear Spin Wave Theory \label{ap-lswt}} 
\begin{figure}
\includegraphics[width=\columnwidth]{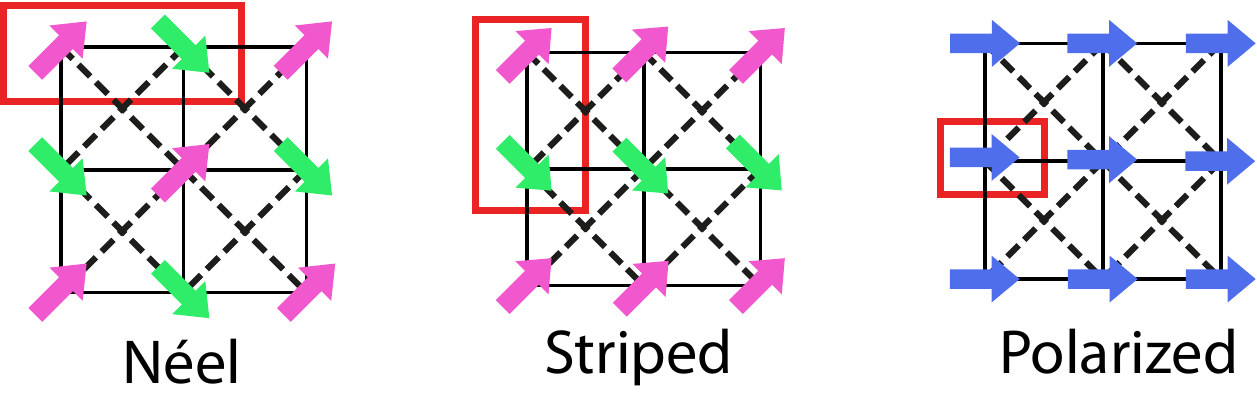}
\caption{(color online) Schematic representations of the classical 
magnetically ordered states around which we consider harmonic quantum 
fluctuations of LSWT. Magnetic unit cells of each classical phase are shown with 
red rectangles.}
\label{ap-fig1}
\end{figure}

We use the three classical reference states shown in Fig.~\ref{ap-fig1} as a background on which harmonic spin waves are considered. Magnetic unit cell of each background state is shown with a red rectangle in Fig.~\ref{ap-fig1}. As a general formalism \cite{Henry:2012}, we define $S_{l,p}$ as the 
{\it p}-th spin ($p=1,\ldots,n$) of the {\it l}-th cell, where $n$ is the number 
of spins in a magnetic unit cell. In the classical limit, an applied transverse 
field $\Gamma$ rotates all spins around the $y$ axis by an angle $\theta$. We 
now introduce a local rotation of spins, as
$S_{l,p} \rightarrow \tilde{S}_{l,p}$, 
in such a way that all three 
classical states shown in Fig.\ref{ap-fig1} map to a simple ferromagnetic state 
in $z$ direction, i.e. $\tilde{S}^{z}_{l,p} = S$ everywhere. Accordingly, we 
define
\begin{eqnarray}
\tilde{S}_{l,p}=\sigma_p R_y(\sigma_p\theta) S_{l,p}
\label{ap-eq3}
\end{eqnarray}
where $\sigma_p=\pm1$ denotes the direction of {\it p}-th spin along the $z$ 
axis, and $R$ is the rotation matrix around $y$ axis by an angle 
$\sigma_p\theta$. Therefore, the following relations between spin components in 
the rotated and non-rotated representations are obtained 
\begin{align}
\nonumber S^z_{l,p}=\sigma_pcos\theta \tilde{S}^z_{l,p} -sin\theta \tilde{S}^x_{l,p},\\
S^x_{l,p}=\sigma_pcos\theta \tilde{S}^x_{l,p}+sin\theta \tilde{S}^z_{l,p}. \label{ap-eq4}
\end{align}
After rewriting the Hamiltonian in terms of new spin operators $\tilde{S}^{z}_{l,p}$ and $\tilde{S}^{x}_{l,p}$, we consider small quantum fluctuations around this general ferromagnetic classical reference state by the following linearized Holstein-Primakoff transformations,
\begin{equation}
\tilde{S}^{z}_{l,p} = S-a_{l,p}^{\dagger}a_{l,p}, \quad \tilde{S}^{x}_{l,p} \approx \sqrt{\dfrac{S}{2}}\left(a_{l,p}^{\dagger}+a_{l,p}\right),	
\label{ap-eq5}
\end{equation}
where $a_{l,p}$ and $a_{l,p}^{\dagger}$ are bosonic operators with well-known 
commutation relations $[a_{l,p},a_{l,p}^{\dagger}]=1$ and 
$[a_{l,p}^{(\dagger)},a_{l,p}^{(\dagger)}]=0$. Hamiltonian is expanded up to  
the quadratic order of bosonic operators, 
\begin{eqnarray}
\mathcal{H}_{LSWT}&&=E_{cl}+\displaystyle\sum_{l,p}{\tilde{h}_p 
a^{\dagger}_{lp}a_{lp}} \\
+\dfrac{1}{8}&&\sin^2\theta\displaystyle \sum_{l,{\bm\delta},p,p'} 
{\tilde{J}({\bm\delta})_{pp'}\left(a^{\dagger}_{l,p}+a_{l,p}\right)
\left(a^{\dagger}_{l',p'}+a_{l',p'}\right)},\nonumber
\label{ap-eq6}
\end{eqnarray}
where linear terms vanish by construction and $\tilde{J}({\bm\delta})$ is an 
$n\times n$ matrix, containing the couplings between spins $p,p'$ of the two 
unit cells $l,l'$ at position ${\bm\delta}$ and 
\begin{eqnarray}
\tilde{h}_p &=& 
-\frac{1}{2}\sigma_{p}\cos^2(\theta)\displaystyle\sum_{{\bm\delta}, 
p'}
\left[\tilde{J}({\bm\delta})\right]_{pp'}\sigma_{p'}+\Gamma\sin\theta.
\label{ap-eq7}
\end{eqnarray}
The momentum space representation is used with the 
following transformations, 
	\begin{eqnarray}
		\nonumber a_{{\bm k},p}&=&\sqrt{\dfrac{n}{N}}\displaystyle\sum_{l}{e^{\mathrm{i}{\bm k}.{\bm 
r}_l}a_{l,p}},\\
	\tilde{J}(\bm k)&=&\displaystyle\sum_{\bm\delta}{e^{-\mathrm{i} {\bm k}.{\bm\delta}}\tilde{J}(\bm \delta)}~.
\label{ap-eq8}
	\end{eqnarray}
Hence, the quadratic Hamiltonian can be written in the following compact form
\begin{eqnarray}
	\nonumber\mathcal{H}_{LSWT} &=& E_{cl}-\dfrac{N}{n}\displaystyle\sum_{p}{\left(\frac{\tilde{h}_p}{2}\right)}
	+\dfrac{1}{2}\displaystyle\sum_{\bm k}{A_{\bm k}^{\dagger}M_{\bm k}A_{\bm k}},\\
	\label{ap-eq9}
\end{eqnarray}
where
\begin{eqnarray}
\nonumber A_{\bm k}^{\dagger} &=& ( a_{{\bm k},1}^{\dagger},\ldots,a_{{\bm k},n}^{\dagger},a_{-{\bm k},1},\ldots,a_{-{\bm k},n} ),\\
\nonumber M_{\bm k}&=&	\begin{pmatrix} 	\tilde{h}+\Delta_{\bm k}&\Delta_{\bm k}\\
														\Delta_{\bm k}&\tilde{h}+\Delta_{\bm k}
							\end{pmatrix},\\
	                        \nonumber [\tilde{h}]_{pp^{\prime}}&=&\tilde{h}_p\delta_{pp^{\prime}},\\
							\Delta_{\bm k}&=&\dfrac{1}{2}\left(\tilde{J}(\bm k) +\tilde{J}(-\bm k)\right).
	\label{ap-eq10}
\end{eqnarray}
Finally, performing an n-mode paraunitary Bogoliubov transformation 
\cite{Colpa:1978}, we obtain the effective diagonal quadratic Hamiltonian given 
by
\begin{equation}
\mathcal{H}_{LSWT} = E_{cl}-
\dfrac{N}{n}\displaystyle\sum_{p}{\left(\frac{\tilde{h}_p}{2}\right)}
+\displaystyle\sum_{\textbf{k},p}{\omega_{\textbf{k},p}
\left(c_{\textbf{k},p}^{
\dagger}c_{\textbf{k},p}+\dfrac{1}{2}\right)},
\label{ap-eq11}
\end{equation}
where $\textbf{k}$ sums over the first Brillouin zone of a lattice constructed 
from the centers of magnetic unit cells of the classical reference states and 
$p$ runs over the spins of a magnetic unit cell, $\tilde{h}_p$ is a correction 
term gained from bosonic commutation relations and $\omega_{\textbf{k},p}$ 
defines the spectrum of quasi-particles with corresponding bosonic creation 
operators $c_{\textbf{k},p}^{\dagger}$. In fact, the eigenmodes 
$\omega_{\textbf{k},p}$ are the eigenvalues of $\Xi M_{\bm k}$, where
$\Xi$ matrix is given by
\begin{eqnarray}
\Xi= \begin{pmatrix}	I_{n}	&	0_{n}\\	0_{n}	&	-I_{n}	
\end{pmatrix} ~.
\label{ap-eq12a}
\end{eqnarray}
Finally, the eigenmodes $\omega_{\textbf{k},p}$ can be
expressed in terms of the eigenvalues $\lambda_{\textbf{k},p}$ of matrix $\Delta_{\bm k}$ in the form 
\begin{eqnarray}
\omega_{\textbf{k},p}=\tilde{h}_p\sqrt{1+2\frac{\lambda_{\textbf{k},p}}{\tilde{h}_p}}.
\label{ap-eq12}
\end{eqnarray}
\section{Excited states in zero field \label{ap-eszf}}
\begin{figure}
\includegraphics[width=0.5\columnwidth]{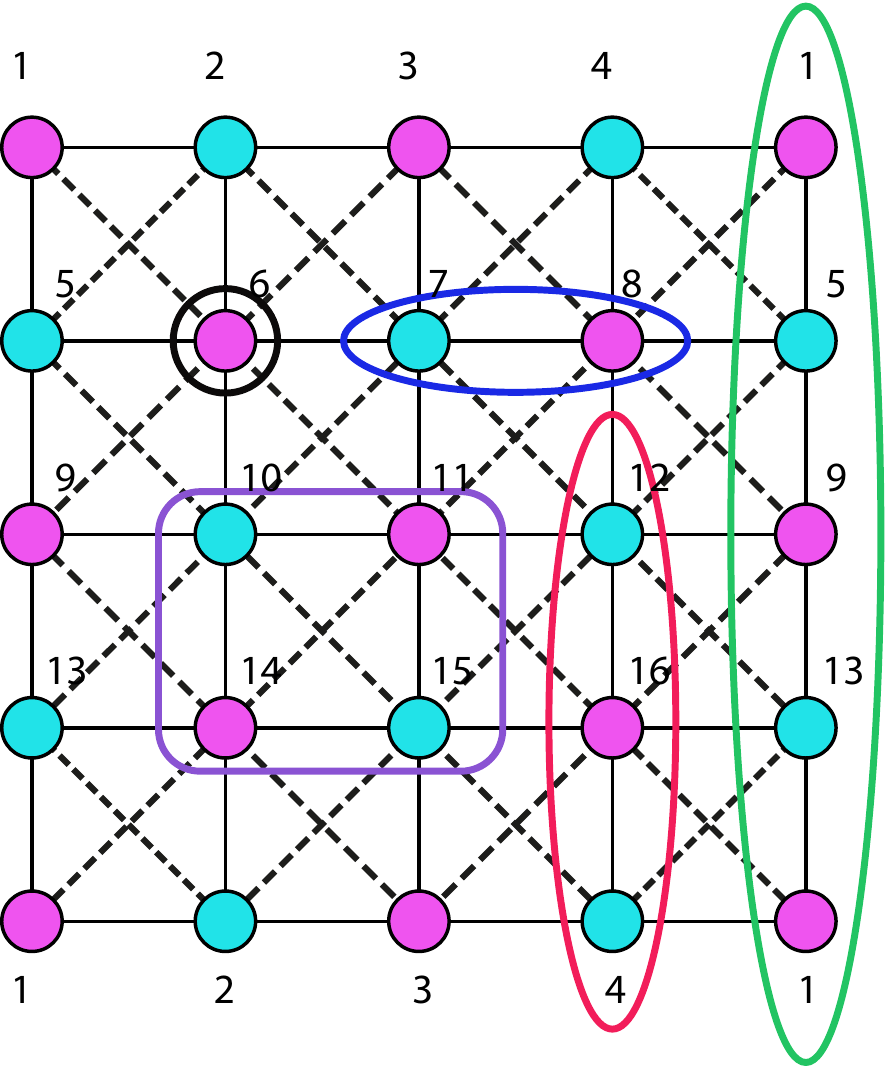}
\caption{(color online) A classical N\'{e}el configuration of Hamiltonian \ref{eq1} at zero field, with periodic boundary conditions on both sides. 
All  $J_1$ bonds are satisfied with this phase
while it is not the case for $J_2$ bonds.
Different kinds of spin excitations 
corresponding to single-spin flip, dimer-flip, trimer-flip, plaquette-flip and 
global-loop flip are shown.}
\label{ap-fig2}
\end{figure}

At the highly frustrated point $J_2/J_1=0.5$ and zero transverse field, the ground 
state of Hamiltonian Eq.~\ref{eq1} is highly degenerate. 
A typical state of this ground space is the 
N\'{e}el state shown in Fig.~\ref{ap-fig2}. The 
lowest-energy excitations might be either a single-spin flip or a 
joint flip of all spins of a specific cluster, which are shown
in Fig.~\ref{ap-fig2}. In a N\'{e}el configuration all nearest-neighbor 
bonds $J_1$ are satisfied, while the next-nearest neighbor bonds $J_2$ are not. Accordingly, flipping one spin will satisfy four $J_2$-bonds,
while dissatisfy four $J_1$-bonds. Hence, the energy cost of a single-spin flip 
excitation is given by $8(J_1-J_2)$, which is equal to $4J_1$ at $J_2/J_1=0.5$. 
Similarly, the energy cost of a dimer-flip, trimer-flip, plaquette flip or every 
finite cluster flip will be $4J_1$. However, a joint flip of all spins on a 
global loop of the lattice (green loop in Fig.~\ref{ap-fig2}) costs 
$4n(J_1-2J_2)$, where $n$ is the number of spins on the global loop. Therefore, 
it implies a zero energy cost at $J_2/J_1=0.5$, which corresponds to the 
transformation of N\'{e}el state to another state of the highly degenerate 
manifold. Therefore, 
the energy cost of a global loop flip is lower than any other finite 
cluster flip, at the highly frustrated point $J_2/J_1=0.5$. 
\begin{figure}
\includegraphics[width=0.6\columnwidth]{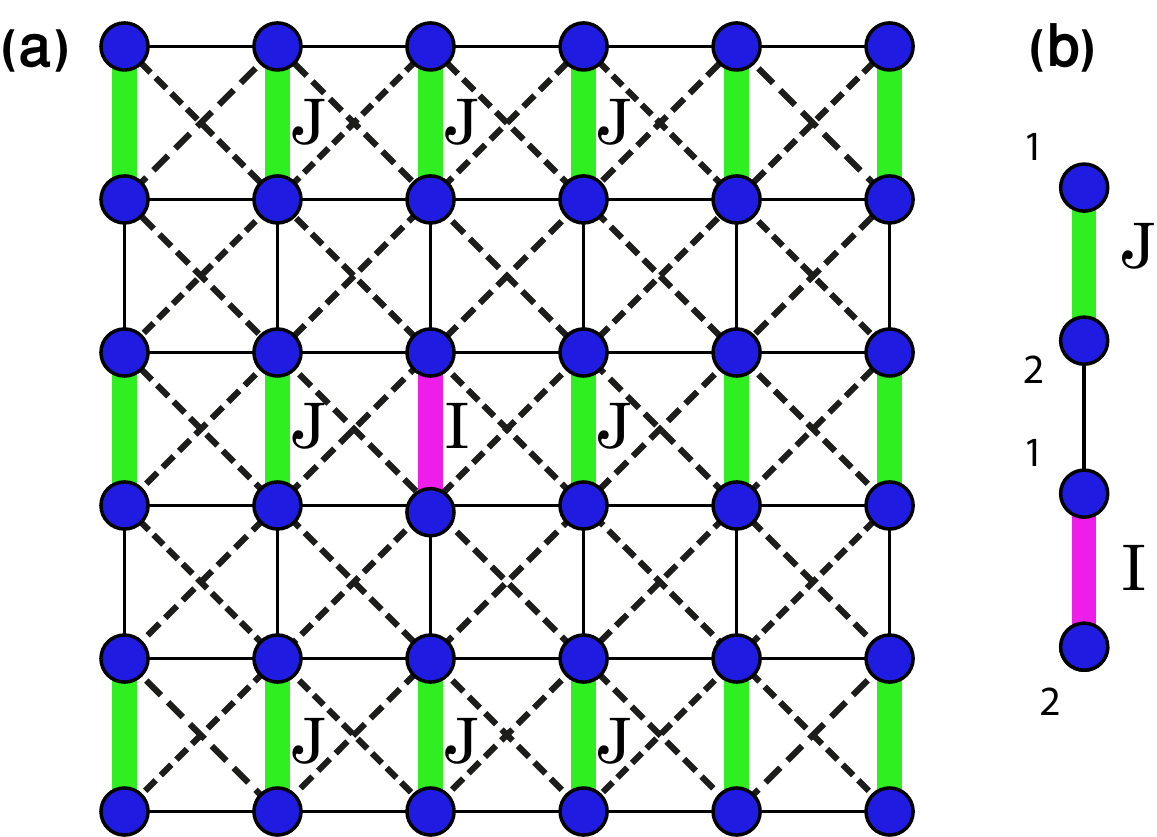}
\caption{(color online) (a) A columnar ordering of dimers as a ground state 
background, used in COA. (b) The interaction between two `nearest-neighbor' 
dimers $I$ and $J$, given by $J_1 S_{1(I)}^z S_{2(J)}^z$.}
\label{ap-fig3}
\end{figure}

\section{Cluster Operator Approach \label{ap-coa}} 
In cluster operator approach (COA), we first consider a perfect multi-spin 
cluster ordering as a ground state background, in which all isolated clusters 
are in their unique ground states. Quantum fluctuations by inter-cluster 
interactions around such ordered reference state give rise to low-lying 
excitations above the perfect cluster ordered background, as a result of 
hybridization of ground state of each cluster to its other excited states, which 
eventuate the zero-point energy correction. In the following, we first propose 
two-spin clusters with columnar orderings shown in Fig.~\ref{ap-fig3}-(a). The 
method for other cluster ordered backgrounds will be similar to this. 

In order to construct an effective theory for the dimer ordered 
background, we rewrite the Hamiltonian \ref{eq1} as 
$\mathcal{H}=\mathcal{H}_0+\mathcal{H}_{int}$, where 
$\mathcal{H}_0=\sum_C{\mathcal{H}}_{C}$ denotes the set of shaded isolated 
dimers shown in Fig.~\ref{ap-fig3} and $\mathcal{H}_{int}$ defines the interaction between them. 
  
\begin{figure}
\includegraphics[width=\columnwidth]{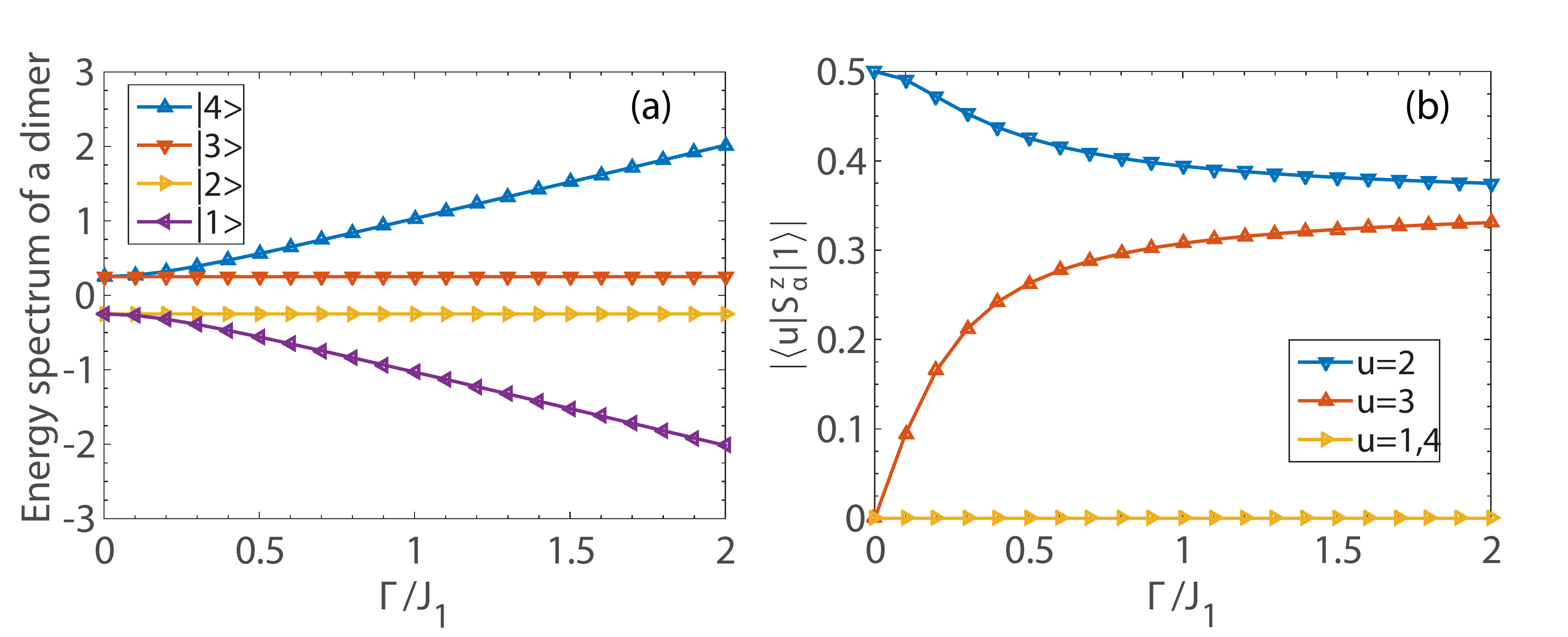}
\caption{(color online) (a) Energy spectrum (in units of $J_1$) of a single dimer versus $\Gamma/J_1$. The bottom line ($\vert1\rangle$) is the unique ground state of the dimer. (b) Transition amplitude $\langle u\vert S_{\alpha}^z\vert1\rangle$ ($\alpha=1,2$), between the ground state $\vert1\rangle$ and four eigenstates $\vert u\rangle$ of a dimer, versus transverse field $\Gamma/J_1$.}
\label{ap-fig4}
\end{figure}
The Hamiltonian of a single dimer is given by
\begin{eqnarray}
\mathcal{H}_{single-dimer}&=&J_1(S_1^z S_2^z)-\Gamma(S_1^x+S_2^x).
\label{ap-eq13}
\end{eqnarray}
The dimer Hamiltonian is diagonalized exactly. The 
energy spectrum as a function of $\Gamma/J_1$ is shown in 
Fig.~\ref{ap-fig4}-(a). It shows a unique ground state $\vert1\rangle$ at 
non-zero transverse field $\Gamma$. 
%Hence, we can define the proposed columnar 
%dimer ordered background (Fig.~\ref{ap-fig3}-(a)) as a state covering the 2D 
%square lattice symmetrically with 2-spin bonds, each of which in its ground 
%state. 
In order to develop an effective theory including inter-dimer 
interactions $\mathcal{H}_{int}$, we first examine the interaction between two 
neighboring dimers. Accordingly, we deduce which excited states of each dimer 
participate in the dynamics of the system when imposing quantum fluctuations 
above the perfect columnar dimer ordered background. Fig.~\ref{ap-fig3}-(a) 
shows that each dimer $I$ interacts with eight neighboring dimers $J$. Let us 
consider interaction between two dimers $I$ and $J$, via a bond $J_1$ between 
spin 1 of dimer $I$ and spin 2 of dimer $J$, shown in the 
Fig.~\ref{ap-fig3}-(b). The interaction, $\mathcal{H}_{IJ}$, is given by
\begin{eqnarray}
\mathcal{H}_{IJ}=J_1S_{1(I)}^zS_{2(J)}^z.
\label{ap-eq14}
\end{eqnarray}
In the absence of this interaction, both dimers I and J are in their unique 
ground states $\vert1\rangle$. Thus, the state of two-dimer system is 
$\vert11\rangle$, i.e. a direct product of single-dimer ground states. Now we 
proceed to turn on the interaction term $\mathcal{H}_{IJ}$ between two dimers, 
as a perturbation. $\mathcal{H}_{IJ}$ does not commute with the isolated 
dimer Hamiltonian, Eq.~\ref{ap-eq13}, which hybridizes the ground 
state of each dimer with its excited states. Accordingly, the matrix 
elements of $\mathcal{H}_{IJ}$ between two direct product states, $\vert 
uv\rangle$ and $\vert11\rangle$, are given by 
\begin{eqnarray}
\langle uv\vert\mathcal{H}_{IJ}\vert11\rangle=J_1\langle u\vert S_{1(I)}^z\vert1\rangle\times\langle v\vert S_{2(J)}^z\vert1\rangle,
\label{ap-eq15}
\end{eqnarray}
where $\vert u\rangle$ and $\vert v\rangle$ are four possible eigenstates of 
dimers I and J, respectively. Fig.~\ref{ap-fig4}-(b) represents the behavior of 
transition amplitude $\langle u\vert S_{\alpha}^z\vert1\rangle$ ($\alpha=1,2$), 
between the ground state $\vert1\rangle$ and four eigenstates $\vert u\rangle$ 
of a single dimer. It shows clearly that for all values of $\Gamma/J_1$, there 
are two excited states $u=2,3$ which dominantly contribute to the dynamics of 
the system as quantum fluctuations. Accordingly, in the following section we 
construct an effective theory for the columnar dimer ordered background via a 
bosonization formalism including only three eigenstates $\vert u\rangle$ 
$(u=1,2,3)$ of each dimer.  

We introduce a bosonization formalism \citep{Sachdev:1990}, similar to what has 
been done in Ref.~\onlinecite{Ganesh:2013,sadrzadeh2015phase} to obtain the 
effective theory. We 
associate a boson to each of the three eigenstates $\vert u\rangle$  
of each dimer $(u=1,2,3)$. In this respect, each eigenstate $\vert u\rangle$ of 
dimer $I$ is created by a boson creation operator $b_{I,u}^\dagger$ acting on 
the vaccum $\vert0\rangle$,
\begin{eqnarray}
\vert u\rangle_I=b_{I,u}^\dagger\vert0\rangle,\    u=1,2,3
\label{ap-eq16}
\end{eqnarray}
where $b_{I,u}^\dagger$ and $b_{I,u}$ are usual bosonic operators, satisfying $[b_{I,u},b_{I,u}^\dagger]=1$ and $[b_{I,u}^{(\dagger)},b_{I,u}^{(\dagger)}]=0$. 
According to the earlier definition, columnar dimer ordered background is 
a Bose-condensate of ground state $\vert u=1\rangle$ bosons, i.e. at a mean 
field level we write 
\begin{eqnarray}
b_{I,1}\equiv b_{I,1}^\dagger\equiv\bar{p},\quad \forall I
\label{ap-eq17}
\end{eqnarray}
where $\bar{p}$ is the condensation amplitude and $\bar{p}^2$ gives the 
probability of a single dimer to be in its ground state. In the absence of 
interaction between dimers, $\bar{p}^2$ is equal to unity. 
However, the existence of inter-dimer interactions 
reduce $\bar{p}^2$ from unity, giving rise to a non-zero occupation of  
excited bosons on single dimers. Nevertheless, to preserve the Hilbert space 
of the effective model, the total occupancy of bosons per dimer should be unity, 
i.e.
\begin{eqnarray}
N_{d} \bar{p}^2+\displaystyle\sum_{I,u=2,3}{b_{I,u}^\dagger b_{I,u}}=N_d,
\label{ap-eq18}
\end{eqnarray}
where $N_d$ is the total number of dimers in Fig.~\ref{ap-fig3}-(a).
Having in mind the Bose-condensation of ground bosons,
the excited bosons are present in very dilute concentrations, 
which lead to neglect the interaction between excited bosons. Hence, we 
only consider the interactions between excited bosons and ground bosons.
In the bosonic language, there are two kinds of inter-dimer interactions participating in the effective Hamiltonian. First, a creation (annihilation) term of excited bosons on neighboring dimers,
\begin{eqnarray}
\nonumber  \vert uv\rangle\langle uv\vert\mathcal{H}_{IJ}\vert11\rangle\langle11\vert&\equiv& d_{uv}^{IJ}\:\bar{p}^2\:b_{I,u}^\dagger b_{J,v}^\dagger\,,\\
  \vert11\rangle\langle11\vert\mathcal{H}_{IJ}\vert uv\rangle\langle uv\vert&\equiv& {d_{uv}^{IJ}}^\dagger\:\bar{p}^2\:b_{I,u} b_{J,v}\,,
\label{ap-eq19}
\end{eqnarray}
and second, a hopping term of excited bosons between neighboring dimers,
\begin{eqnarray}
\nonumber  \vert u1\rangle\langle u1\vert\mathcal{H}_{IJ}\vert1v\rangle\langle1v\vert&\equiv& h_{uv}^{IJ}\:\bar{p}^2\:b_{I,u}^\dagger b_{J,v}\,,\\
  \vert1v\rangle\langle1v\vert\mathcal{H}_{IJ}\vert u1\rangle\langle u1\vert&\equiv& {h_{uv}^{IJ}}^\dagger\:\bar{p}^2\:b_{I,u} b_{J,v}^\dagger\,,
\label{ap-eq20}
\end{eqnarray}
where coefficients $d_{uv}^{IJ}=\langle uv\vert\mathcal{H}_{IJ}\vert11\rangle$ 
and $h_{uv}^{IJ}=\langle u1\vert\mathcal{H}_{IJ}\vert1v\rangle$ are creation and 
hopping amplitudes, respectively. On the other hand, according to 
Fig.~\ref{ap-fig4}-(b), the values of terms like $\langle1\vert 
S_{\alpha}^z\vert1\rangle$ are zero, which rules out $\mathcal{O}(\bar{p}^3)$ and  
$\mathcal{O}(\bar{p}^4)$ terms in the effective Hamiltonian. Those terms independent of $\bar{p}$ and $\mathcal{O}(\bar{p})$ can be ignored due to 
neglecting interaction between excited bosons.

Based on the above arguments, the effective Hamiltonian for the columnar dimer 
ordered background can now be written in a quadratic bosonic form, 
\begin{eqnarray}
\nonumber \mathcal{H}&=&N_d\bar{p}^2\epsilon_{1}+\sum_{I}\sum_{u}\epsilon_ub_{I,u}^\dagger b_{I,u}\\
 \nonumber &&-\mu\Big[N_d\bar{p}^2 +\sum_{I,u}b_{I,u}^\dagger b_{I,u}-N_d\Big]\\
\nonumber &&+\bar{p}^2\sum_{\langle I,J\rangle}\sum_{u,v} \Big[(d_{uv}^{IJ})\:b_{I,u}^\dagger b_{J,v}^\dagger+(h_{uv}^{IJ})\: b_{I,u}^\dagger b_{J,v}+ H.c.\Big].\\
\label{ap-eq21}
\end{eqnarray}
The first line includes intra-dimer terms, where the index $I$ sums over all 
isolated dimers in Fig.~\ref{ap-fig3}-(a) and $u$ sums over the two dominant 
excited states ($u=2,3$) of each dimer with corresponding eigen energies 
$\epsilon_u$. The second line enforces 
the Hilbert space constraint, Eq.~\ref{ap-eq18}, via a chemical potential 
$\mu$. The third line involves inter-dimer terms, where $u,v=2,3$ are the two 
excited bosons of neighboring dimers $I$ and $J$, respectively. It is remarkable 
that the eigenstates of a single dimer Hamiltonian, according to 
Eq.~\ref{ap-eq13}, have  Z$_2$ symmetry. It implies that all of the bosonic 
states participating in the effective Hamiltonian keep this symmetry. Therefore, 
the Z$_2$ symmetry of the original Hamiltonian Eq.~\ref{eq1} is respected in our 
effective theory of Eq.~\ref{ap-eq21}.

In order to diagonalize the effective Hamiltonian, we first 
go to the momentum space by introducing the Fourier transform of the bosonic 
operators and interactions,
\begin{eqnarray}
b_{\textbf{k},u} = \frac{1}{\sqrt{N_D}}\sum_{\textbf{k}} b_{I,u}e^{-i\textbf{k}.\textbf{r}_I},\hspace{1em} \nonumber H_{\textbf{k}}=\sum_{\textbf{k}} H_{IJ} e^{i\textbf{k}.(\textbf{r}_J-\textbf{r}_I)},\\
\label{ap-eq22}
\end{eqnarray}
where $\textbf{k}$ sums over the first Brillouin zone of a rectangular lattice formed by the centers of columnar dimers of Fig.~\ref{ap-fig3}-(a). Finally, having done a paraunitary Bogoliubov transformation \cite{Colpa:1978}, the effective Hamiltonian takes the diagonal form 
\begin{eqnarray}
\nonumber \mathcal{H}&=& N_D\mu+ N_D\bar{p}^2(\epsilon_1- \mu) -\frac{1}{2}N_D\sum_{u=2,3}(\epsilon_u-\mu)
\\  
&&+\sum_{\textbf{k}} \sum_{\nu=1}^{2}\big(\frac{1}{2} + \gamma^\dagger_{\nu, \textbf{k}}\gamma_{\nu, \textbf{k}}\big)\Omega_{\nu, \textbf{k}}(\mu,\bar{p}),
\label{ap-eq23}
\end{eqnarray}
where $\Omega_{\nu,\textbf{k}}$ gives the eigenmodes of the effective model, 
corresponding to the bosonic excitations $\gamma^\dagger_{\nu, \textbf{k}}$ 
around the columnar dimer ordered background. These excitation modes cause the 
zero-point energy corrections for the ground state.
Two parameters 
$\bar{p}$ and $\mu$ are specified self consistently, by solving the following 
equations
\begin{eqnarray}
\frac{\partial \langle \mathcal{H} \rangle}{\partial \mu} = 0, 
\hspace{1em} \frac{\partial \langle \mathcal{H} \rangle}{\partial \bar{p}} = 0. 
\label{ap-eq24}
\end{eqnarray}

It should be mentioned that the above procedure is essentially a numerical 
task for clusters larger than four sites. For instance, we have to take 
256 states into account for $\ell=10$ cluster to consider non-vanishing 
transition amplitudes.

\section{Tree Tensor Network \label{ap-ttn}}
In tensor network formalism, we could represent each quantum many-body state in 
terms of local tensors connected through geometric structures 
\cite{Verstraete-Matrix-2008}. The geometric structures are 
determined by global properties appeared in the system, such as entanglement and 
or correlations. In principle, faithful tensor network states should have the 
ability to reproduce all global features apeared in the system. For instance, low-lying excited states of local 
Hamiltonians respect area law, stating bipartite entanglement entropy (of 
subsystem) scales by common boundary of two partitions instead of volume 
\cite{Eisert:2010}. Furthermore, two-point correlation function for gapped and 
gapless phases respectively decay exponentially and algebraically, as distance 
between two partitions increases. So, the reliable tensor network states are 
ones that are cleverly designed to fulfill such behavior, specially pattern of 
entanglement and correlation are of important ones. 

\begin{figure}
\includegraphics[width=\columnwidth]{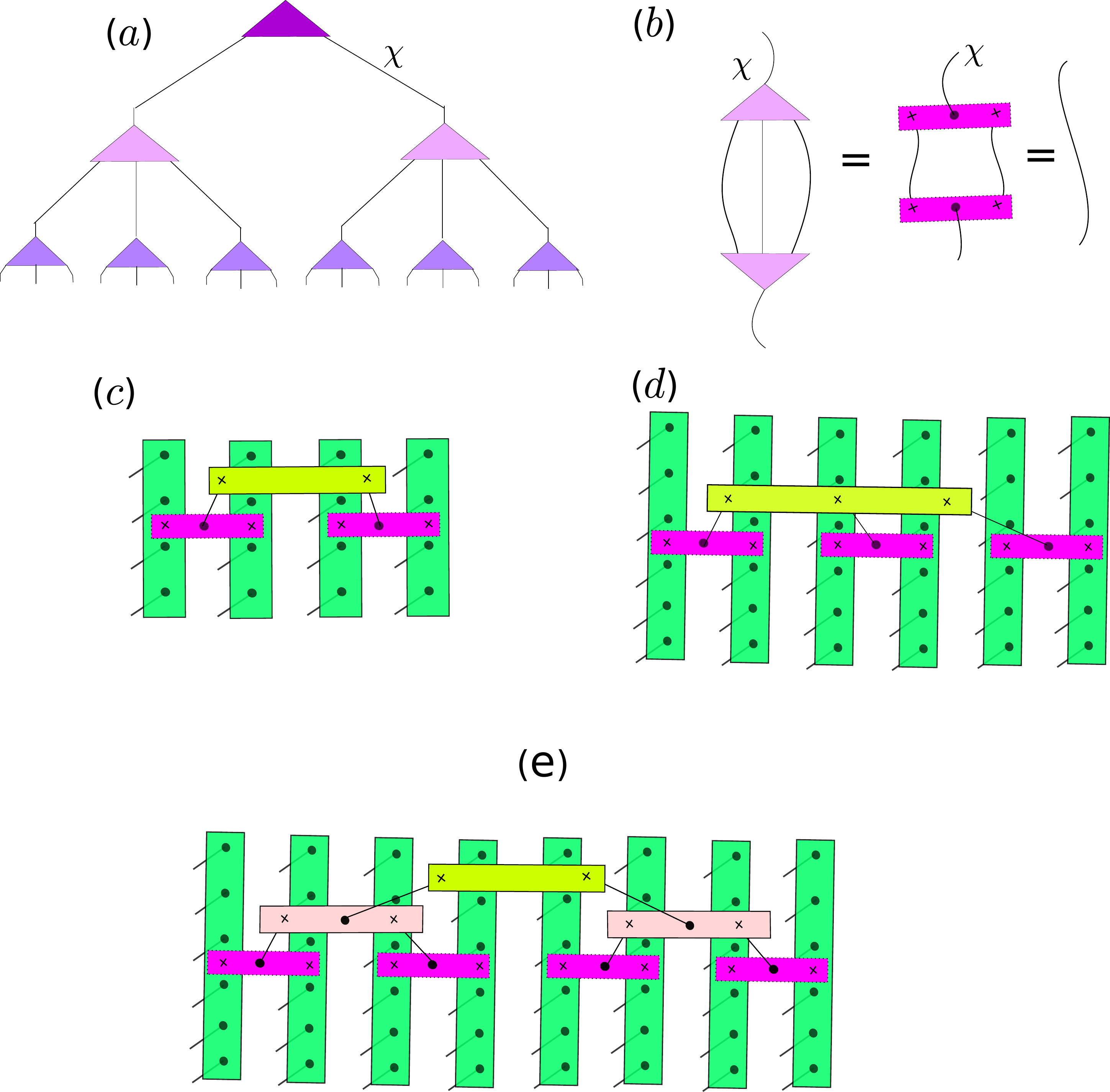}
\caption{(color online) Graphical representation of TTN. $(a)$ 1D TTN, $(b)$ 
isometric constraint, 2D TTN for $(c)$ $4\times4$, $(d)$ $6 \times 6$ and $(e)$ 
$6 \times 8$ latices.}
\label{ap-fig5}
\end{figure}

TTN is a class of tensor network states inspired by renormalization 
group methodology, i.e. Wilsona's and Kadanoff's earlier works 
\cite{Efrati:2014}. TTN states are represented in terms of local 
isometric tensors (see Fig.~\ref{ap-fig5}-(a, b)) forming a tree-like geometric 
graph. Such tree-like structures have some numerical/conceptual advantages, 
making TTN as a powerful numerical toolbox: $(i)$ different types of 
optimization method could be simply applied \cite{Gerster:2014, Evenbly:2009}, 
$(ii)$ reduces time/memory cost of the algorithm, and $(iii)$ reproduces 
algebraic behavior of correlation function and so on. 
However, 2D TTNs are suitable only for small clusters, since it violets area 
law---as it occurs for matrix product states. In 
Fig.~\ref{ap-fig5}-(a), we have shown a $3$-layer 1D TTN composing of isometric 
triangular tensors. The triangular tensors play the role of RG steps, mapping 3 
spins into a superspin with effective bond-dimension $\chi$. At each layer, they 
are the same, exploiting translational invariant symmetry. One could easily 
generalize 1D TTN to 2D cases, as we have shown them for $4\times 4$, $6\times 
6$ and $6\times 8$ clusters, respectively, in Fig.~\ref{ap-fig5}-(c, d, e). We 
exactly utilize these 2D TTNs in our simulations.

We follow Ref.~\onlinecite{Tagliacozzo:2009} to perform optimization algorithm: 
the main idea is to take a specific local isomeric tensor---fixing the other 
tensors---as variational parameters and then obtain variational ground-state 
energy, so that it becomes minimum. By repeating this process over all other 
tensors, TTN state would hopefully converge to real ground state. Bond-dimension 
$\chi$ is our control parameter determining accuracy of algorithm---it is 
obvious for $\chi \rightarrow \infty $, the result would be exact. Time and 
memory cost of optimization processing respectively scale by 
$\mathcal{O}(\chi^{4})$ and $\mathcal{O}(\chi^{3})$. Calculating expectation 
value of local operators, (nearest neighbor) correlation function and 
variational energy have also the same cost. In our calculation, we consider 
clusters up to $8\times 6$ spins, and also do finite-$\chi$ scaling to obtain 
more accurate result \cite{Pollmann:2008}. We use the following equation to 
obtain our final data  
\begin{equation}
<\widehat{\mathcal{O}}>_{\chi}=<\widehat{\mathcal{O}}>_{\infty} + 
\frac{A_{0}}{\chi^{\theta}},
\end{equation}
where $<\widehat{\mathcal{O}}>$ stand for expectation value of operators, 
$A_{0}$ and $\theta$ are two constants---determined by the best fitting 
methods. 
Note $<\widehat{\mathcal{O}}>_{\infty}$ is the quantity which is reported 
throughout the paper. We take $\chi \sim 400$ so that error in variational 
ground-state energy, in the worst cases (critical point), is of order 
$10^{-4}$. 

%%%%%%%%%%%%%%%%%%%%%%%%%%%%%%%%%%%%%%%%%%%%%%%%%%%%%%%

%%%%%%%%%%%%%%%%%%%%%%%%%%%%%%%%%%%%
\end{document}